\documentclass[manuscript, anonymous=false, screen]{acmart}

\AtBeginDocument{%
  }

\setcopyright{none} 
\acmConference[IMF]{extended version of the 15th International Conference on IT Security Incident Management \& IT Forensics}{2026}{}

\usepackage{float} 
\usepackage{multirow} 

\usepackage{enumitem}

\newenvironment{packed_enum}{
\begin{enumerate}[leftmargin=0.4cm]
  \setlength{\itemsep}{1pt}
  \setlength{\parskip}{0pt}
  \setlength{\parsep}{0pt}
}{\end{enumerate}}

\newenvironment{packed_item}{
\begin{itemize}[leftmargin=0.4cm]
  \setlength{\itemsep}{1pt}
  \setlength{\parskip}{0pt}
  \setlength{\parsep}{0pt}
}{\end{itemize}}

\usepackage{amsthm}

\usepackage{csquotes}
\usepackage[table]{xcolor}

\usepackage{caption}
\usepackage{subcaption}

\renewcommand{\labelenumi}{\alph{enumi})}
\renewcommand\footnotetextcopyrightpermission[1]{}
\begin{document}

\title[Exploring Usability and Legal Practice]{Exploring Usability and Legal Practice: Insights from German Judicial Users of Digital Forensics}

\author{Tobias Hoppmann}
\orcid{0009-0000-0971-6636}
\affiliation{%
  \institution{Friedrich-Alexander-Universität Erlangen-Nürnberg}
  \city{Erlangen}
  \country{Germany}
}
\email{tobias.hoppmann@fau.de}

\author{Leona Lassak}
\orcid{0000-0001-8309-3211}
\affiliation{%
  \institution{Ruhr University Bochum}
  \city{Bochum}
  \country{Germany}}
\email{leona.lassak@rub.de}

\author{M. Angela Sasse}
\orcid{0000-0003-1823-5505}
\affiliation{%
  \institution{Ruhr University Bochum}
  \city{Bochum}
  \country{Germany}}
\email{martina.sasse@rub.de}

\author{Zinaida Benenson}
\orcid{0009-0006-7158-0219}
\affiliation{%
  \institution{Friedrich-Alexander-Universität Erlangen-Nürnberg}
  \city{Erlangen}
  \country{Germany}}
\email{zinaida.benenson@fau.de}

\renewcommand{\shortauthors}{Hoppmann et al.}

\begin{abstract}
    Digital forensics has become an integral part of modern criminal proceedings, yet its effective integration remains challenging because of increasing data volumes, evolving technologies, and complex interactions between technical and legal stakeholders. Although prior work has focused primarily on digital forensic tools and methods, its broader procedural and organizational context has received limited attention. Building on emerging perspectives inspired by usability research and human-centered security, we conceptualize digital forensics as part of a socio-technical system within criminal proceedings. We consequently investigate this perspective through a survey of 101 practitioners from the judiciary of the German federal state of North Rhine-Westphalia, including public prosecutors, judges, and digital forensic experts. The results indicate a strong demand for improved integration of digital forensics into workflows, enhanced cross-domain communication, and a closer alignment of stakeholder expectations. 
    They also uncover great potential for the improvement of digital forensics usability, e.g., through stronger interdisciplinary cooperation, easier and faster access to  evidential data and results, or improvement of stakeholder training and education.
    
\end{abstract}


\maketitle

\emph{This manuscript is based on the paper accepted for presentation at the 15th International Conference on IT Security Incident Management \& IT Forensics (IMF 2026)\footnote{\url{https://imf-conference.org/}}.
It provides the complete survey questionnaire in Appendix~\ref{app1}.}

\section{Introduction}
\label{sec:intro}

Digital evidence has become a central component of most criminal proceedings~\citep{EU.2025}. The discipline concerned with the identification, acquisition, analysis, and presentation of this evidence is digital forensics~(DF), which has become an established element of modern criminal investigations. Despite increasing efforts to strengthen this field in recent years, DF faces growing challenges~\citep{Garfinkel.2010,Garfinkel.2022,Hargreaves.2024}. The widespread integration of information technology and its complexities into daily life means criminals become more technologically sophisticated, and leverage new tools and domains to enable highly organized and specialized forms of criminal activity.

While organized crime is offering tools that criminals with little technical expertise can use, law enforcement and DF are lagging behind. At the same time, the rapid provision of digital evidence often plays a key role in criminal proceedings, e.g. when reviewing the justification for detention of subjects within established time limits. 

Discussions about usability aspects of DF have largely focused on the technical aspects of specific tools~\citep{Bennett.2008, Hibshi.2011}. More recent work extends this perspective by drawing on human-centered security to examine the broader socio-technical context of criminal proceedings and their previously neglected user communities~\citep{Hoppmann.2026, Warner.2026}.

Our study explores usability issues in DF with stakeholders in criminal proceedings, to identify practical challenges in the handling and integration of DF into the process. We focus on the three basic  usability criteria effectiveness, efficiency, and satisfaction (see Section~\ref{sub: Usability}) --- including collaboration, organizational structures, communication, training, and investigator well-being.
Our study participants were 101 members of the judiciary of the German state of North Rhine-Westphalia, representing most of the stakeholders involved in DF in criminal proceedings: DF experts, investigators, prosecutors, and judges. In addition, this sample also provides insight into the perspective of judicial authorities such as courts and public prosecution offices, which is often overlooked.

This exploratory study was guided by the following research questions:

\begin{packed_item}
    \item \emph{RQ1:} How do perceptions of digital forensics differ between stakeholder groups involved in criminal proceedings?
    \item \emph{RQ2:} Which recurring challenges in usability of digital forensics are perceived by stakeholders as relevant from their perspective?
\end{packed_item}

\paragraph{Contributions} This paper makes the following contributions:
\begin{packed_item}
    \item First, we identify relevant areas, information sources, and challenges in DF that are linked to usability and are part of the broader socio-technical context of criminal proceedings.
    \item We then present exploratory survey findings from 101 practitioners involved in criminal proceedings, including judges, public prosecutors, DF expert, and investigators, providing insights into challenges related to the handling and integration of digital evidence and DF.
    \item Finally, we discuss the results and their implications for future research in the field of usability in DF.
\end{packed_item}

\paragraph{Roadmap}
Section~\ref{sec:new_background} provides an overview of the fundamentals of usability, the current state of usability research in DF, and some specific features of the German criminal justice system. The methodology and development of the survey questions are presented in Section~\ref{sec: Methods}. Section~\ref{sec:results} presents the results of the survey. The findings are discussed in Section~\ref{sec:discussion}, before the paper is summarized in Section~\ref{sec:conclusion}.


\section{Background}
\label{sec:new_background}

 We begin considering the key concepts of usability, and how they apply within the context of DF. We then provide an overview of selected research areas in DF and highlight recurring challenges related to the three usability criteria effectiveness, efficiency, and satisfaction, which are explained below. Finally, we outline key characteristics of German criminal proceedings that are relevant to the survey. 

\subsection{Usability}\label{sub: Usability}
The concept of usability is described by the ISO/IEC 9241-11 standard~\citep{ISO.2018} and is defined as \textit{``the extent to which a system, product, or service can be used by specified users to achieve specified goals with effectiveness, efficiency, and satisfaction in a specified context of use.''} The standard defines effectiveness as the accuracy and completeness with which users achieve goals, and efficiency as the resources expended (e.g., time, effort, cost) relative to outcomes. Satisfaction reflects the perceptions and cognitive, physical, and emotional responses of users, shaped by their needs and expectations, and influences both behavior and performance.

The above definition also introduces the concept ``context of use'', which refers to the environment in which the systems, products, or services under consideration are used. The context of use includes users, goals, tasks, resources, and the surrounding technical, physical, social, and organizational conditions.

\subsection{Usability in Digital Forensics}\label{sub: Usability in Digital Forensics}
 
A small number of studies have addressed usability in DF, with most focusing on individual tools. 
For example, the ``Autopsy Forensic Browser'' was evaluated by~\citet{Bennett.2008} using cognitive walkthrough and heuristic evaluation methods. Their results highlight significant usability challenges, including complex technical terminology, inadequate error handling, and workflow limitations caused by the underlying system architecture.~\citet{Hibshi.2011} examine multiple forensic tools through interviews and surveys, and relate these findings to the cognitive responses of users. They show that participants need tools that require less cognitive effort, support collaborative work environments, and have high interoperability. 


A qualitative study by~\citet{Nouh.2019} underlines the need for improved coordination and information sharing in cybercrime investigations. The authors criticize poor usability of current tools which lack interoperability, data visualization functions, and adequate user support through human in the loop design concepts, and advocate for better training, workflow adaptation, and greater engagement from the human-computer interaction~(HCI) community.

From a practical perspective,~\citet{Pollitt.2013} emphasizes that DF tools are increasingly unable to keep pace with the rapidly evolving technological and investigative demands, and that technical proficiency alone has never been sufficient to ensure success in DF. Instead, he highlights interdisciplinary collaboration among the different stakeholders, and organizational factors like the integration of DF into the surrounding investigative context, as a key determinant of its effectiveness.

Although not explicitly focused on usability, a study by~\citet{Flory.2016} provides insights on the adoption and use of DF within law enforcement, available expertise and training, and perceptions regarding the relevant capabilities of prosecutors’ offices and courts in Indiana, US. The study identifies a lack of funding, insufficient training, and inadequate use of academic resources to support investigations as potential hurdles in the area of DF capabilities within law enforcement.

The extended perspective on usability of DF and its application in criminal proceedings are further developed and formalized by~\citet{Hoppmann.2026}. They propose a model that classifies usability issues along human, organizational, and technical aspects, and introduce a concept and a model of usability of DF in criminal proceedings based on ISO/IEC 9241-11~\citep{ISO.2018}. 
In subsequent research,~\citet{Hoppmann.2026b} show that integrating usability into DF does not undermine its scientific and legal  principles~\citep{Dewald.2015}.

As shown in the field of usable security~\citep{Adams.1999,Whitten.1999,sasse2001transforming,Beautement.2008}, poor recognition of users’ capabilities, needs, tasks, and goals can lead to misunderstandings, friction, and reduced collaboration. This is reflected~\citet{Bossler.2012} in the perception of the responsibility for cybercrime investigations by patrol officers. Approximately 73\% of the officers surveyed believed that digital evidence should be handled by specialized DF units. But a shortage of trained professionals creates a bottleneck, and to speed up proceedings, the securing and review of digital evidence is often delegated to police officers with only basic training.~\citet{WilsonKovacs.2020}, The core tasks also include rapid preliminary analysis - technical triage - to assess the value of digital evidence being searched.~\citet{Warner.2026} highlight the low usability of such triage approaches, which can lead to a sense of learned helplessness and the development of problematic workarounds.~\citet{Pollitt.2013} adds that triage measures are often directives issued by executives without consulting the departments concerned, including police officers and DF units.

In conclusion, the international large-scale DFPulse expert survey by~\citet{Hargreaves.2024} depicts the challenges faced by DF experts. These include organizational challenges such as high workload, insufficient staff, tight budgets, inadequate training, and significant backlogs. According to~\citet{Reason.2008}, these often sequential organizational, structural, and sometimes cultural weaknesses can be assumed to be latent failures. Together with individual failures, these can cause accidents or loss of control in situations.

In Appendix~\ref{sub: Related Issues in Digital Forensics}, we present an extended literature overview of related work in DF, divided into the three characteristics of usability according to the ISO/IEC 9241-11 standard: effectiveness, efficiency and satisfaction. This extended literature analysis was used to precisely identify and define the survey questions.

\subsection{Distinctive Features of German Criminal Proceedings}\label{sub: Specialties in German Criminal Proceedings}

Given the significant differences in legal principles and frameworks across jurisdictions, this study focuses on German criminal proceedings. We explain key distinctive features of German criminal proceedings that are relevant for this study.

\paragraph{Formal Stages of Criminal Proceedings}
Under the German Code of Criminal Procedure~(GCCP), the formal stages of criminal proceedings differ from other legal frameworks or generalized descriptions such as those proposed by~\citet{Seepma.2020}. 
Notably, the GCCP does not include a separate detection phase, in which the police operate independently of the public prosecution office~(PPO), with situational exceptions, e.g., for imminent danger. 
The primary responsibility for maintaining public safety lies with the police authorities, while the PPO is responsible for criminal prosecution.
The investigation phase, led or overseen by the PPO, marks the formal beginning of the proceedings. Although the PPO is legally responsible for directing the investigation and should be involved from the outset, the police typically carry out most investigative tasks on behalf of the PPO.  The PPO can issue investigative directives, but does not exercise control over police resources or broader operational priorities. Public prosecutors~(PPs) also determine whether to bring charges, initiating a preliminary court review. If the court accepts the indictment, the main trial phase begins, during which evidence is presented and the courts gain ownership of the case. A subsequent enforcement phase, also managed by the PPO, may follow the court's verdict ~\cite{.20241107, Gless.2021}.

\paragraph{The Role of the Courts}

Section 244 (2) of the GCCP~\citep{.20241107} establishes the principle that the court must investigate the facts of the case on its own initiative. This means that judges actively direct the court proceedings, determine which evidence is to be taken, actively question witnesses or have them questioned, and bear responsibility for establishing the truth. In particular, judges may also order further investigations.

\paragraph{Digital Evidence at Court}
Due to the regulations of the GCCP, digital evidence must be presented in a form suitable for court proceedings. The GCCP declares that the evidence must fit one of the four strict categories of admissible evidence for court proceedings: \textit{(I)}~witness evidence, \textit{(II)}~expert evidence, \textit{(III)}~documentary evidence, or \textit{(IV)}~evidence by personal inspection (cf.~\cite{Gless.2021,.20241107} sec.~244~to~256). Since German law does not provide for a separate category for digital evidence, it must be adapted to fit into one of the four strict categories. Typically, digital evidence primarily plays a supportive role due to the necessary adaptation process. This allows the application of strict rules of evidence without requiring additional regulations specifically for digital evidence. Instead, guidelines for conducting DF, such as those issued by the German Federal Office for Information Security~(BSI), are relied upon~\cite{BSIBundesamtfurSicherheitinderInformationstechnik.2011,Gless.2021}. 

\paragraph{Investigative Authorities and Digital Forensics Units}
In the German criminal justice system, in addition to the police, the Federal Customs Service and independent experts, there are other agencies that operate their own DF units. For example, many federal states have established focal public prosecution offices~(focal-PPOs), which in some cases employ their own DF experts. In addition, DF experts are also found within tax investigation divisions.


\section{Method}\label{sec: Survey Design}\label{sec: Methods}

The objective of this study is to investigate usability-related challenges in DF within criminal proceedings. The survey focuses on the stakeholders' perspectives on how usability influences workflows, collaboration, and the effective handling of digital evidence and DF across organizational boundaries. The study examines multiple dimensions of DF practice, capturing human, technical, and organizational aspects. Usability considerations serve as a cross-cutting perspective that informs several dimensions, including challenges, collaboration, and training (see Appendix~\ref{sub: Related Issues in Digital Forensics}). 

\subsection{Concept and Derivation of Survey Items}\label{sub: Conceptual Dimensions of the Survey}\label{sub: Derivation of Survey Items}

\begin{figure*}[hb]
    \centering
    \includegraphics[width=1\linewidth]{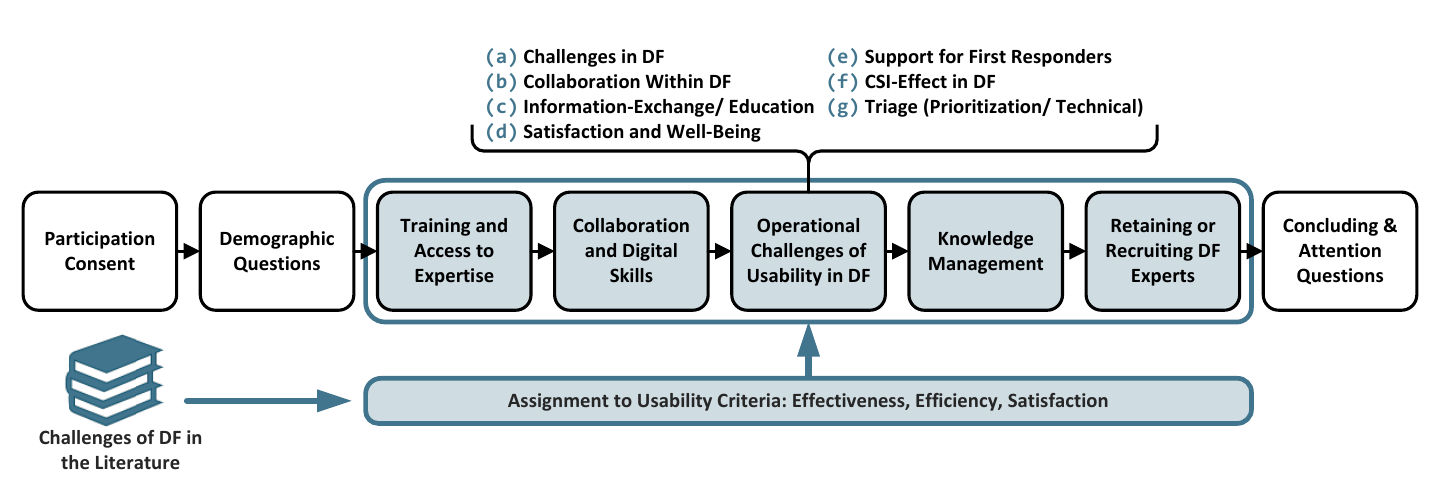}
    \caption{Overview of the structure and order of the survey’s question groups, including the further breakdown of the “Operational Challenges of Usability in DF” group. The rationale for the colored groups is described in Sections~\ref{quest: Training and Access to Expertise} to~\ref{quest: Retaining or Recruiting DF Experts}. The illustration also highlights the origin of these question groups from the literature described in Section~\ref{sub: Usability in Digital Forensics} and Appendix~\ref{sub: Related Issues in Digital Forensics}, as well as their connection to usability through the three usability criteria described in Section~\ref{sub: Usability}.}
    \Description{Overview of the order of the survey’s eight question groups—from participant consent to the attention question. The five core groups—``Training and Access to Expertise,'' ``Collaboration and Digital Skills,'' ``Operational Challenges of Usability in DF,'' ``Knowledge Management,'' and ``Retaining or Recruiting DF Experts''—are highlighted in color. The group ``Operational Challenges of Usability in DF'' is further subdivided into ``Challenges in DF,'' ``Collaboration Within DF,'' ``Information Exchange/Education,'' ``Satisfaction and Well-Being,'' ``Support for First Responders,'' ``CSI Effect in DF,'' and ``Triage.''}
    \label{fig:SurveyDesign}
\end{figure*}

The survey is divided into eight groups of questions that reflect key issues in DF usability in criminal proceedings, visualized in Figure~\ref{fig:SurveyDesign}. Following consent to participate and demographic questions, the survey addresses training and access to expertise, collaboration, and digital competencies, as well as a central section on operational challenges. Finally, the survey concludes with questions about the future perceived relevance of usability in DF and the quality of the responses. 

Given the diversity of the backgrounds of the respondents, and to keep completion time of survey questions manageable, answers were given via Likert-scale items. The survey questions can be viewed in Appendix~\ref{app1}.

\subsubsection{Training and Access to DF Expertise}\label{quest: Demographics, Knowledge and Availability of Experts}\label{quest: Training and Access to Expertise}
This group includes questions about whether the respondents’ organizations employ DF experts. If so, the number and professional background of these experts are recorded, and if not, the reasons for not having them are explored. In addition, participants are asked if they participated in DF-related training sessions over the past five years, and the content of those sessions (technical or crime-related topics).

\subsubsection{Collaboration and Digital Skills}\label{quest: cooperation and competence}
Extending Flory's study~\citep{Flory.2016}, which focused on prosecutors and judges, the survey covers a broader set of stakeholders of DF in criminal proceedings, including courts, PPOs, focal PPOs, police, tax investigators, academia, and the private business sector. The initial question of this group captures both the collaboration patterns and the perceived DF-related competencies.

Participants were also asked to rate the  competence of their organization in handling digital evidence. This line of questions was also applied to additional stakeholder groups and dimensions, including the effectiveness of investigations, resource availability, and satisfaction. Possible answers include the respondent's own organization and the organizations with which the respondent has previously indicated a collaboration.

\subsubsection{Operational Challenges of Usability in DF}\label{quest: Operational Challenges and Usability Issues}
In addition, we included 15 Likert-scale questions that address the challenges identified in Sections~\ref{sub: Usability in Digital Forensics} and Appendix~\ref{sub: Related Issues in Digital Forensics}, grouped into seven thematic sub-areas. The order of the questions is randomized.

\begin{packed_enum}
    
    \item Three questions examined \emph{Challenges in DF}, including the impact of evidence backlogs on investigations~\citep{Hargreaves.2024, Pollitt.2013, WilsonKovacs.2020}, the need for established scientific procedures and methods~\citep{Palmer.2001}, and the potential value of distinct categories for digital evidence within legal frameworks~\citep{Gless.2021}.

    \item \emph{Collaboration} questions assessed the role of interdisciplinary teams and decentralization~\citep{Hansen.2017, Casey.2019}, contrasting perspectives on centralization~\citep{Nouh.2019}, the demand for Digital-Forensics-as-a-Service~(DFaaS) solutions~\citep{Baar.2014}, and issues related to the reluctance and responsibility of first responders~\citep{Bossler.2012, Warner.2026, WilsonKovacs.2020}.

    \item \emph{Information Sharing} was addressed by examining its impact on DF efficiency~\citep{Nouh.2019, WilsonKovacs.2020} and the need for training and knowledge transfer between domains~\citep{Casey.2021, Hansen.2017}, including integration into professional education.

    \item \emph{Recognition and Well-Being} were captured through two questions on work conditions and feedback, reflecting their impact on effectiveness and satisfaction~\citep{Kelty.2021, Strickland.2023}.

    \item We included a question on \emph{Technical Support for Search Teams and First Responders}, motivated by the high risk of errors during early evidence collection~\citep{Casey.2021}.

    \item One question examined the \emph{“CSI Effect”}, i.e., the potential influence of forensic science portrayals in media on expectations about forensics~\citep{Cole2009}. This item aimed to assess whether such perceptions extend to DF.
    
    \item Two questions addressed \emph{Triage}, distinguishing between general and technical triage~\citep{Horsman.2022}, focusing on the efficiency and the effectiveness, including risks such as false negatives and irreversible errors. Since triage appears to be an area of significant friction between different user groups and at the organizational level~\citep{Pollitt.2013, Warner.2026, WilsonKovacs.2020}, the responses to this are of considerable relevance for usability research.
    
\end{packed_enum}

\subsubsection{Knowledge Management in DF}\label{quest: Knowledge Management in DF}
Silos of DF-related knowledge and inter-organizational collaboration pose significant challenges in the domain of DF. Knowledge Management Systems~(KMSs) affect all three usability criteria and intersect with the human, technical, and organizational aspects of usability in DF~\citep{Hoppmann.2026}. Thus, KMSs represent a paradigmatic case for usability in DF.

To operationalize this topic, we draw on previous work by~\citet{Casey.2021} on the mature and field-tested KMS KIES. These questions were shown if participants had previously confirmed the need for KMS. Based on~\citet{Casey.2021}, we derived ten system functions and seven usage aspects, which were assessed through two multiple-choice questions that addressed their perceived importance. 

\subsubsection{Retaining or Recruiting DF Experts}\label{quest: Retaining or Recruiting DF Experts}

We included questions on the perceptions of the participants about recruitment challenges, both in general and within DF. Although these items were originally intended for a broader law enforcement context and are constrained by the structure of state-level collective bargaining frameworks, they are reported here due to their potential relevance for DF.

The inclusion of these questions was also motivated by the ongoing community discussion on well-being in among DF experts. Preliminary findings from an international well-being study\footnote{International Well-Being Study: What The Early Data Shows, published 17 February 2026 by Forensic Focus, \url{https://www.forensicfocus.com/news/international-well-being-study-what-the-early-data-shows/}} conducted by Northumbria University indicate that 36\% of the DF experts surveyed plan to leave the profession within the next 12 months, with an additional 20\% uncertain. Although these results are not yet final, they highlight the urgency of retaining and paying attention to the well-being of DF experts. 

Two additional questions examine potential conflicts between the IT sector and law enforcement, which could contribute to greater skepticism between the two fields. One such conflict was identified in the specifically German debate over the so-called “hacker paragraphs.” These provisions can quickly put legitimate IT security research in the crosshairs of law enforcement agencies. These questions aim to clarify whether such an effect is considered likely, as it could hinder the recruitment of new IT experts~\citep{Freiling.2023, Golla.2023, Nolde.2023}.


\subsection{Usability as an Overarching Concern}\label{sub: The Role of Usability as an Overarching Concern}

Although the survey covers a broad range of topics, item selection follows a consistent underlying principle: alignment with core usability criteria effectiveness, efficiency, and satisfaction, introduced in Section~\ref{sub: Usability}. 

The survey questions were systematically mapped to one or more usability criteria, guided by a structured categorization of issues identified in the literature in Section~\ref{sub: Usability in Digital Forensics} and Appendix~\ref{sub: Related Issues in Digital Forensics}.

This perspective also informed the broader structure of the survey. For example, KMS are treated as a distinct topic, as they inherently span multiple usability criteria, including effectiveness (e.g., quality of knowledge transfer), efficiency (e.g., access to information), and satisfaction, while also supporting collaboration. Similarly, the section on cooperation and competencies captures usability-related aspects of inter-organizational interaction, including perceived effectiveness, efficiency, and satisfaction.

Other sections, such as personnel and organizational culture, further reflect usability aspects, particularly in light of emerging discussions on well-being in DF. In general, this cross-cutting perspective provides a unifying framework for interpreting survey results across otherwise diverse directions.


\subsection{Instrument Refinement and Validation}\label{sub: Instrument Refinement and Validation}

The survey was revised iteratively to ensure clarity, relevance, and suitability for the target group. The initial drafts of the survey were reviewed by different experts with practical experience in DF, police, and judiciary, which led to adjustments in the language, structure, and order of the questions. Particular attention was paid to improving clarity and reducing cognitive load, especially given the diverse professional backgrounds of the respondents and the length of the survey.

\subsection{Ethical Considerations} \label{sub: Ethical Considerations}

To improve the acceptance of the study within the judiciary and law enforcement, the survey did not require case-specific information and minimized psychologically sensitive content. In addition, a draft questionnaire was shared with the relevant ministries to solicit feedback on possible concerns. No concerns were raised by any of the ministries consulted.

Especially, highly sensitive topics such as child sexual abuse material~(CSAM) were deliberately excluded. Although such cases are part of the practice of DF, the primary focus of this study is on the usability of DF. While selected findings from related research (see Section~\ref{sec:new_background} and Appendix~\ref{sub: Related Issues in Digital Forensics}) are considered, the emphasis is on usability aspects rather than CSAM-specific issues. We recognize the importance of usability in reducing the burden associated with CSAM investigations, but we are also aware of the special responsibility that comes with conducting research on this topic. 

Since no faculty ethics board was in place at the time (it is in the process of being established), the study adhered to the principles of the Menlo Report~\citep{Menlo.12}, which is a recognized ethics guideline for computer security research. The main identified risk concerned participant identification. Therefore, no direct identification information (e.g.,~names, years of birth) or gender data was collected. The latter was not included because the sample was expected to consist of small groups with an unbalanced gender ratio, and the identification of individual participants based on personal characteristics had to be ruled out as much as possible. All responses were stored under pseudonymous IDs to ensure confidentiality.

Participation was voluntary and anonymous, with no personal, network, or sensitive data collection. An introductory text explained the study’s objectives, estimated completion time, data protection measures, and contact details. No compensation was offered. Participants confirmed their informed consent before starting the survey and could pause or terminate the survey at any time, with the option to resume within one week. 

The online survey was implemented with Qualtrics XM\footnote{\url{https://www.qualtrics.com}} in an own instance of the software. The data were stored in an access-protected and encrypted environment to which only project members had access.

\subsection{Recruitment}\label{sub: Recruitment}

The recruitment was organized through the ministries of the federal state of North Rhine-Westphalia in Germany. 
Official requests to participate were sent to the responsible ministries representing the judiciary and investigative organizations involved in criminal proceedings. 
Ultimately, the judiciary participated in the study.
Since the judiciary agencies cover the relevant stages and key stakeholders in criminal proceedings, the recruitment was successful. To maximize participation, invitations and additional flyers with QR codes to access the survey were distributed through the internal communication channels of the judiciary.

\subsection{Data Analysis}\label{sub: Data Analysis}
The survey used different question formats depending on the research objective. These included single-choice and multiple-choice questions with varying numbers of answer options, along with questions on a Likert scale. The Likert scale questions typically used a five-category response format, consisting of two opposing poles and a neutral option in the middle, to capture variations in the perceptions and opinions of participants. Descriptive statistical analyzes were conducted for all types of questions, while inferential statistical methods were only applied where suitable.

In the four questions about organizational competencies assessed in the handling of digital evidence in Section~\ref{quest: cooperation and competence}, we varied the standard Likert scale. To provide users with a familiar interface for evaluation, responses were collected using a nine-point Likert-type scale ranging from 0.5 to 5 in increments of 0.5, conceptually aligned with star-based rating systems commonly used in practice (e.g., consumer review platforms). The scale starts at 0.5, as this ensured a response for every option, thus eliminating the tenth response option of a standardized ten-point Likert scale.

In Section~\ref{sub: Collaboration and Digital Skills}, we conducted a statistical significance test on the results of the subgroups of PPs with and without access to internal DF units on these questions.
Group differences were analyzed using the non-parametric Mann-Whitney $U$ test due to the ordinal nature of Likert-scale data. Since non-parametric statistical methods based on rank ordering were applied, the specific scaling does not affect the validity of the analysis. We controlled for multiple comparisons by adjusting the $p$-values using the False Discovery Rate (FDR, Benjamini-Hochberg method~\citep{benjamini1995controlling}). 

Statistical significance was assessed using the FDR-corrected $p$-value (*$p < 0.05$, **$p < 0.01$, ***$p < 0.001$). The substantive relevance of the effects was assessed using the correlation coefficient $r$ according to~\citet{Cohen.1988} and classified as small ($r\geq0.1$), moderate ($r\geq0.3$), or large ($r\geq0.5$). A result is considered substantial only if it is statically significant and has at least a moderate effect size.

\subsection{Demographics}\label{sub: Demographics}

We received responses from a total of 101 participants in the judicial sector. Table~\ref{tab: demo} summarizes the key demographic characteristics of the sample. The study includes 15 judges involved in criminal proceedings, 46~prosecutors, and 10 executives. The last group consists of eight senior prosecutors in leadership positions and two executives from the administration and IT. Seven DF experts participated, which is noteworthy, as they are typically members of focal PPOs and are much less common in the judiciary than, e.g., in the police. The survey also reached groups of investigators with other areas of expertise, administrative staff (including IT), and other groups. In addition to group affiliation, the survey collected data on age groups, employment status, and the highest level of education.

\begin{table}[htbp]
  \centering
  \caption{Overview of key demographic variables from the survey, stratified by judges, public prosecutors (PPs), digital forensic experts (DF), investigators (Invest.), executives (Exec.), administration/IT staff (Adm./IT), and others.}
    \begin{tabular}{p{15em}rrrrrrrr}
    \textbf{Demographics} & \multicolumn{1}{l}{\textbf{Judges}} & \multicolumn{1}{l}{\cellcolor[rgb]{ .949,  .949,  .949}\textbf{PPs}} & \multicolumn{1}{l}{\textbf{DF}} & \multicolumn{1}{l}{\cellcolor[rgb]{ .949,  .949,  .949}\textbf{Invest.}} & \multicolumn{1}{l}{\textbf{Exec.}} & \multicolumn{1}{l}{\cellcolor[rgb]{ .949,  .949,  .949}\textbf{Adm./IT}} & \multicolumn{1}{l}{\textbf{Misc.}} & \multicolumn{1}{l}{\cellcolor[rgb]{ .749,  .749,  .749}\textbf{Total}} \\
    \midrule
    \textbf{Number of participants} & 15 & \cellcolor[rgb]{ .949,  .949,  .949}46 & 7 & \cellcolor[rgb]{ .949,  .949,  .949}12 & 10 & \cellcolor[rgb]{ .949,  .949,  .949}8 & 3 & \cellcolor[rgb]{ .749,  .749,  .749}\textbf{101} \\
    \rowcolor[rgb]{ .949,  .949,  .949} \textbf{\% of participants (on total)} & 14.9\% & \cellcolor[rgb]{ .851,  .851,  .851}45.5\% & 6.9\% & \cellcolor[rgb]{ .851,  .851,  .851}11.9\% & 9.9\% & \cellcolor[rgb]{ .851,  .851,  .851}7.9\% & 3.0\% & \cellcolor[rgb]{ .651,  .651,  .651}\textbf{100\%} \\
    \midrule
    \multicolumn{9}{l}{\textbf{Age cohorts (n = 101)}} \\
    \midrule
    \textbf{Age from 18 to 29} & 0 & \cellcolor[rgb]{ .949,  .949,  .949}0 & 1 & \cellcolor[rgb]{ .949,  .949,  .949}0 & 1 & \cellcolor[rgb]{ .949,  .949,  .949}2 & 1 & \cellcolor[rgb]{ .749,  .749,  .749}\textbf{5} \\
    \rowcolor[rgb]{ .949,  .949,  .949} \textbf{Age from 30 to 39} & 6 & \cellcolor[rgb]{ .851,  .851,  .851}25 & 3 & \cellcolor[rgb]{ .851,  .851,  .851}2 & 0 & \cellcolor[rgb]{ .851,  .851,  .851}4 & 1 & \cellcolor[rgb]{ .651,  .651,  .651}\textbf{41} \\
    \textbf{Age from 40 to 49} & 4 & \cellcolor[rgb]{ .949,  .949,  .949}15 & 0 & \cellcolor[rgb]{ .949,  .949,  .949}6 & 2 & \cellcolor[rgb]{ .949,  .949,  .949}1 & 0 & \cellcolor[rgb]{ .749,  .749,  .749}\textbf{28} \\
    \rowcolor[rgb]{ .949,  .949,  .949} \textbf{Age from 50 to 59} & 5 & \cellcolor[rgb]{ .851,  .851,  .851}5 & 1 & \cellcolor[rgb]{ .851,  .851,  .851}3 & 5 & \cellcolor[rgb]{ .851,  .851,  .851}1 & 0 & \cellcolor[rgb]{ .651,  .651,  .651}\textbf{20} \\
    \textbf{Age from 60 to 69} & 0 & \cellcolor[rgb]{ .949,  .949,  .949}1 & 2 & \cellcolor[rgb]{ .949,  .949,  .949}1 & 2 & \cellcolor[rgb]{ .949,  .949,  .949}0 & 1 & \cellcolor[rgb]{ .749,  .749,  .749}\textbf{7} \\
    \midrule
    \multicolumn{9}{l}{\textbf{Organizational affiliation (n = 101)}} \\
    \midrule
    \rowcolor[rgb]{ .949,  .949,  .949} \textbf{Courts} & 15 & \cellcolor[rgb]{ .851,  .851,  .851}0 & 0 & \cellcolor[rgb]{ .851,  .851,  .851}0 & 0 & \cellcolor[rgb]{ .851,  .851,  .851}0 & 0 & \cellcolor[rgb]{ .651,  .651,  .651}\textbf{15} \\
    \textbf{(Focal-) Public Prosecution Office} & 0 & \cellcolor[rgb]{ .949,  .949,  .949}46 & 7 & \cellcolor[rgb]{ .949,  .949,  .949}12 & 8 & \cellcolor[rgb]{ .949,  .949,  .949}3 & 2 & \cellcolor[rgb]{ .749,  .749,  .749}\textbf{78} \\
    \rowcolor[rgb]{ .949,  .949,  .949} \textbf{Administration/ IT} & 0 & \cellcolor[rgb]{ .851,  .851,  .851}0 & 0 & \cellcolor[rgb]{ .851,  .851,  .851}0 & 2 & \cellcolor[rgb]{ .851,  .851,  .851}5 & 1 & \cellcolor[rgb]{ .651,  .651,  .651}\textbf{8} \\
    \midrule
    \multicolumn{9}{l}{\textbf{Employment status (n = 101)}} \\
    \midrule
    \rowcolor[rgb]{ .949,  .949,  .949} \textbf{Employee} & 1 & \cellcolor[rgb]{ .851,  .851,  .851}0 & 7 & \cellcolor[rgb]{ .851,  .851,  .851}6 & 1 & \cellcolor[rgb]{ .851,  .851,  .851}6 & 3 & \cellcolor[rgb]{ .651,  .651,  .651}\textbf{24} \\
    \textbf{Civil servant} & 14 & \cellcolor[rgb]{ .949,  .949,  .949}46 & 0 & \cellcolor[rgb]{ .949,  .949,  .949}6 & 9 & \cellcolor[rgb]{ .949,  .949,  .949}2 & 0 & \cellcolor[rgb]{ .749,  .749,  .749}\textbf{77} \\
    \midrule
    \multicolumn{9}{l}{\textbf{Classification of the position - pay and grade (n = 101)}} \\
    \midrule
    \rowcolor[rgb]{ .949,  .949,  .949} \textbf{Higher Service} & 15 & \cellcolor[rgb]{ .851,  .851,  .851}44 & 7 & \cellcolor[rgb]{ .851,  .851,  .851}6 & 10 & \cellcolor[rgb]{ .851,  .851,  .851}1 & 0 & \cellcolor[rgb]{ .651,  .651,  .651}\textbf{83} \\
    \textbf{Upper Intermediate Service} & 0 & \cellcolor[rgb]{ .949,  .949,  .949}2 & 0 & \cellcolor[rgb]{ .949,  .949,  .949}6 & 0 & \cellcolor[rgb]{ .949,  .949,  .949}5 & 0 & \cellcolor[rgb]{ .749,  .749,  .749}\textbf{13} \\
    \rowcolor[rgb]{ .949,  .949,  .949} \textbf{Intermediate Service} & 0 & \cellcolor[rgb]{ .851,  .851,  .851}0 & 0 & \cellcolor[rgb]{ .851,  .851,  .851}0 & 0 & \cellcolor[rgb]{ .851,  .851,  .851}1 & 2 & \cellcolor[rgb]{ .651,  .651,  .651}\textbf{3} \\
    \textbf{Other pay scale} & 0 & \cellcolor[rgb]{ .949,  .949,  .949}0 & 0 & \cellcolor[rgb]{ .949,  .949,  .949}0 & 0 & \cellcolor[rgb]{ .949,  .949,  .949}0 & 1 & \cellcolor[rgb]{ .749,  .749,  .749}\textbf{1} \\
    \rowcolor[rgb]{ .949,  .949,  .949} \textbf{Not specified} & 0 & \cellcolor[rgb]{ .851,  .851,  .851}0 & 0 & \cellcolor[rgb]{ .851,  .851,  .851}0 & 0 & \cellcolor[rgb]{ .851,  .851,  .851}1 & 0 & \cellcolor[rgb]{ .651,  .651,  .651}\textbf{1} \\
    \midrule
    \multicolumn{9}{l}{\textbf{Highest qualification achieved  (n = 101)}} \\
    \midrule
    \multicolumn{1}{p{15em}}{\textbf{Master’s degree} or higher/ equiv. } & 15 & \cellcolor[rgb]{ .949,  .949,  .949}41 & 5 & \cellcolor[rgb]{ .949,  .949,  .949}7 & 9 & \cellcolor[rgb]{ .949,  .949,  .949}2 & 1 & \cellcolor[rgb]{ .749,  .749,  .749}\textbf{80} \\
    \rowcolor[rgb]{ .949,  .949,  .949} \multicolumn{1}{p{15em}}{\textbf{Bachelor’s degree} or equiv.} & 0 & \cellcolor[rgb]{ .851,  .851,  .851}5 & 2 & \cellcolor[rgb]{ .851,  .851,  .851}2 & 1 & \cellcolor[rgb]{ .851,  .851,  .851}1 & 0 & \cellcolor[rgb]{ .651,  .651,  .651}\textbf{11} \\
    \multicolumn{1}{p{15em}}{\textbf{Certified Specialist} or equiv.} & 0 & \cellcolor[rgb]{ .949,  .949,  .949}0 & 0 & \cellcolor[rgb]{ .949,  .949,  .949}2 & 0 & \cellcolor[rgb]{ .949,  .949,  .949}1 & 0 & \cellcolor[rgb]{ .749,  .749,  .749}\textbf{3} \\
    \rowcolor[rgb]{ .949,  .949,  .949} \textbf{Vocational training} & 0 & \cellcolor[rgb]{ .851,  .851,  .851}0 & 0 & \cellcolor[rgb]{ .851,  .851,  .851}1 & 0 & \cellcolor[rgb]{ .851,  .851,  .851}4 & 2 & \cellcolor[rgb]{ .651,  .651,  .651}\textbf{7} \\
    \end{tabular}%
  \label{tab: demo}%
\end{table}%

\section{Results}\label{sec:results}

This section presents the survey findings, structured according to the thematic question groups introduced in Section~\ref{sub: Derivation of Survey Items}. 

\subsection{Training and Access to DF Expertise}\label{sub: Content of Training Courses in Digital Forensics and Investigations}

As shown in Table~\ref{tab: training}, 38 participants (40\% of non-DF experts) reported access to internal DF expertise. Including experts from DF (n~=~45), the unit size most reported was four to six experts (73.3\%).

Among the 38 participants who reported no or uncertain access, the reasons for the absence of DF experts were lack of funding (17), other reasons (12), no perceived need (9), reliance on external experts (7), lack of applicants~(3) and sufficient internal expertise~(1). 

In general, 44.6\% of the participants reported having attended DF-related training in the past five years, while~37.6\% confirmed access to internal DF expertise.

To assess training content, six subject areas were defined—three technical (DF, digital evidence, cybercrime/IT security) and three legal/criminal (investigative, legal and phenomenological aspects)---with multiple responses allowed. As shown in Table~\ref{tab: training}, technical topics predominate, a pattern that remains largely unchanged when excluding DF experts, administration/IT, and other groups. However, the results in Table~\ref{tab: training} should be interpreted with caution: 
For example, although all judges who confirmed participation in DF-related training reported a 100\% coverage of the topic ``digital evidence'', only 20\% of all judges who participated in the survey (3 of 15) had received such training.

\begin{table*}[htbp]
  \centering
  \caption{Overview of responses on training content among the 45 participants who received training in cybercrime, digital evidence or DF within the past five years (multiple responses allowed).}
\begin{tabular}{p{16em}rrrrrrrr}
\multicolumn{1}{l}{\textbf{DF Training and Internal Expertise}} & \textbf{Judges}  & \multicolumn{1}{l}{\cellcolor[rgb]{ .949,  .949,  .949}\textbf{PPs}} & \multicolumn{1}{l}{\textbf{  DF  }} & \multicolumn{1}{l}{\cellcolor[rgb]{ .949,  .949,  .949}\textbf{Invest.}} & \multicolumn{1}{l}{\textbf{Exec.}} & \multicolumn{1}{l}{\cellcolor[rgb]{ .949,  .949,  .949}\textbf{Adm./IT}} & \multicolumn{1}{l}{\textbf{Misc.}} & \multicolumn{1}{l}{\cellcolor[rgb]{ .749,  .749,  .749}\textbf{Total}} \\
\midrule
\multicolumn{9}{l}{\textbf{Internal availability of DF expertise (n = 94 - without DF experts)}} \\
\midrule
\multicolumn{1}{l}{\textbf{Yes}} & 0 & \cellcolor[rgb]{ .949,  .949,  .949}17 & \multicolumn{1}{c}{-} & \cellcolor[rgb]{ .949,  .949,  .949}11 & 5 & \cellcolor[rgb]{ .949,  .949,  .949}3 & 2 & \cellcolor[rgb]{ .749,  .749,  .749}\textbf{38} \\
\rowcolor[rgb]{ .949,  .949,  .949} \multicolumn{1}{l}{\textbf{No}} & 9 & \cellcolor[rgb]{ .851,  .851,  .851}16 & \multicolumn{1}{c}{-} & \cellcolor[rgb]{ .851,  .851,  .851}1 & 4 & \cellcolor[rgb]{ .851,  .851,  .851}4 & 1 & \cellcolor[rgb]{ .651,  .651,  .651}\textbf{35} \\
\multicolumn{1}{l}{\textbf{Don't know}} & 6 & \cellcolor[rgb]{ .949,  .949,  .949}13 & \multicolumn{1}{c}{-} & \cellcolor[rgb]{ .949,  .949,  .949}0 & 1 & \cellcolor[rgb]{ .949,  .949,  .949}1 & 0 & \cellcolor[rgb]{ .749,  .749,  .749}\textbf{21} \\
\midrule
\multicolumn{9}{l}{\textbf{Number of DF experts (only if “availability” is selected or in group of DF expert - n = 45)}} \\
\midrule
\rowcolor[rgb]{ .949,  .949,  .949} \multicolumn{1}{l}{\textbf{One (1)}} & 0 & \cellcolor[rgb]{ .851,  .851,  .851}1 & 1 & \cellcolor[rgb]{ .851,  .851,  .851}0 & 0 & \cellcolor[rgb]{ .851,  .851,  .851}0 & 0 & \cellcolor[rgb]{ .651,  .651,  .651}\textbf{2} \\
\multicolumn{1}{l}{\textbf{Two to three (2-3)}} & 0 & \cellcolor[rgb]{ .949,  .949,  .949}1 & 0 & \cellcolor[rgb]{ .949,  .949,  .949}1 & 1 & \cellcolor[rgb]{ .949,  .949,  .949}2 & 0 & \cellcolor[rgb]{ .749,  .749,  .749}\textbf{5} \\
\rowcolor[rgb]{ .949,  .949,  .949} \multicolumn{1}{l}{\textbf{Four to six (4-6)}} & 0 & \cellcolor[rgb]{ .851,  .851,  .851}12 & 6 & \cellcolor[rgb]{ .851,  .851,  .851}10 & 4 & \cellcolor[rgb]{ .851,  .851,  .851}1 & 0 & \cellcolor[rgb]{ .651,  .651,  .651}\textbf{33} \\
\multicolumn{1}{l}{\textbf{Don't know}} & 0 & \cellcolor[rgb]{ .949,  .949,  .949}3 & 0 & \cellcolor[rgb]{ .949,  .949,  .949}0 & 0 & \cellcolor[rgb]{ .949,  .949,  .949}0 & 2 & \cellcolor[rgb]{ .749,  .749,  .749}\textbf{5} \\
\midrule
\multicolumn{9}{l}{\textbf{Participation in DF-related training courses within the last 5 years (n = 101)}} \\
\midrule
\multicolumn{1}{l}{\textbf{Yes}} & 3 & \cellcolor[rgb]{ .949,  .949,  .949}16 & 6 & \cellcolor[rgb]{ .949,  .949,  .949}8 & 6 & \cellcolor[rgb]{ .949,  .949,  .949}5 & 1 & \cellcolor[rgb]{ .749,  .749,  .749}\textbf{45} \\
\rowcolor[rgb]{ .949,  .949,  .949} \multicolumn{1}{l}{\textbf{No}} & 12 & \cellcolor[rgb]{ .851,  .851,  .851}30 & 1 & \cellcolor[rgb]{ .851,  .851,  .851}4 & 4 & \cellcolor[rgb]{ .851,  .851,  .851}3 & 2 & \cellcolor[rgb]{ .651,  .651,  .651}\textbf{56} \\
\midrule
\multicolumn{9}{p{40.85em}}{\textbf{Content of DF-related training courses (n = 45, multiple choice)}} \\
\midrule
\rowcolor[rgb]{ .949,  .949,  .949} \textbf{Digital Evidence} & 3 & \cellcolor[rgb]{ .851,  .851,  .851}10 & 5 & \cellcolor[rgb]{ .851,  .851,  .851}5 & 2 & \cellcolor[rgb]{ .851,  .851,  .851}1 & 1 & \cellcolor[rgb]{ .651,  .651,  .651}\textbf{27} \\
\textbf{Technical Cybercrime/IT Security} & 2 & \cellcolor[rgb]{ .949,  .949,  .949}5 & 6 & \cellcolor[rgb]{ .949,  .949,  .949}3 & 4 & \cellcolor[rgb]{ .949,  .949,  .949}5 & 1 & \cellcolor[rgb]{ .749,  .749,  .749}\textbf{26} \\
\rowcolor[rgb]{ .949,  .949,  .949} \textbf{Digital Forensics} & 0 & \cellcolor[rgb]{ .851,  .851,  .851}8 & 5 & \cellcolor[rgb]{ .851,  .851,  .851}6 & 1 & \cellcolor[rgb]{ .851,  .851,  .851}2 & 1 & \cellcolor[rgb]{ .651,  .651,  .651}\textbf{23} \\
\textbf{Criminalistic Aspects} & 0 & \cellcolor[rgb]{ .949,  .949,  .949}9 & 2 & \cellcolor[rgb]{ .949,  .949,  .949}5 & 3 & \cellcolor[rgb]{ .949,  .949,  .949}2 & 0 & \cellcolor[rgb]{ .749,  .749,  .749}\textbf{21} \\
\rowcolor[rgb]{ .949,  .949,  .949} \textbf{Legal Aspects, including IT Law} & 2 & \cellcolor[rgb]{ .851,  .851,  .851}6 & 3 & \cellcolor[rgb]{ .851,  .851,  .851}1 & 4 & \cellcolor[rgb]{ .851,  .851,  .851}2 & 0 & \cellcolor[rgb]{ .651,  .651,  .651}\textbf{18} \\
\textbf{Phenomenology/ Manifestations} & 0 & \cellcolor[rgb]{ .949,  .949,  .949}9 & 2 & \cellcolor[rgb]{ .949,  .949,  .949}2 & 3 & \cellcolor[rgb]{ .949,  .949,  .949}1 & 0 & \cellcolor[rgb]{ .749,  .749,  .749}\textbf{17} \\
\end{tabular}%
  \label{tab: training}%
\end{table*}%

\subsection{Collaboration and Digital Skills}\label{sub: Collaboration and Digital Skills}

\begin{figure*}[b]
    \centering
    \includegraphics[width=0.7\linewidth]{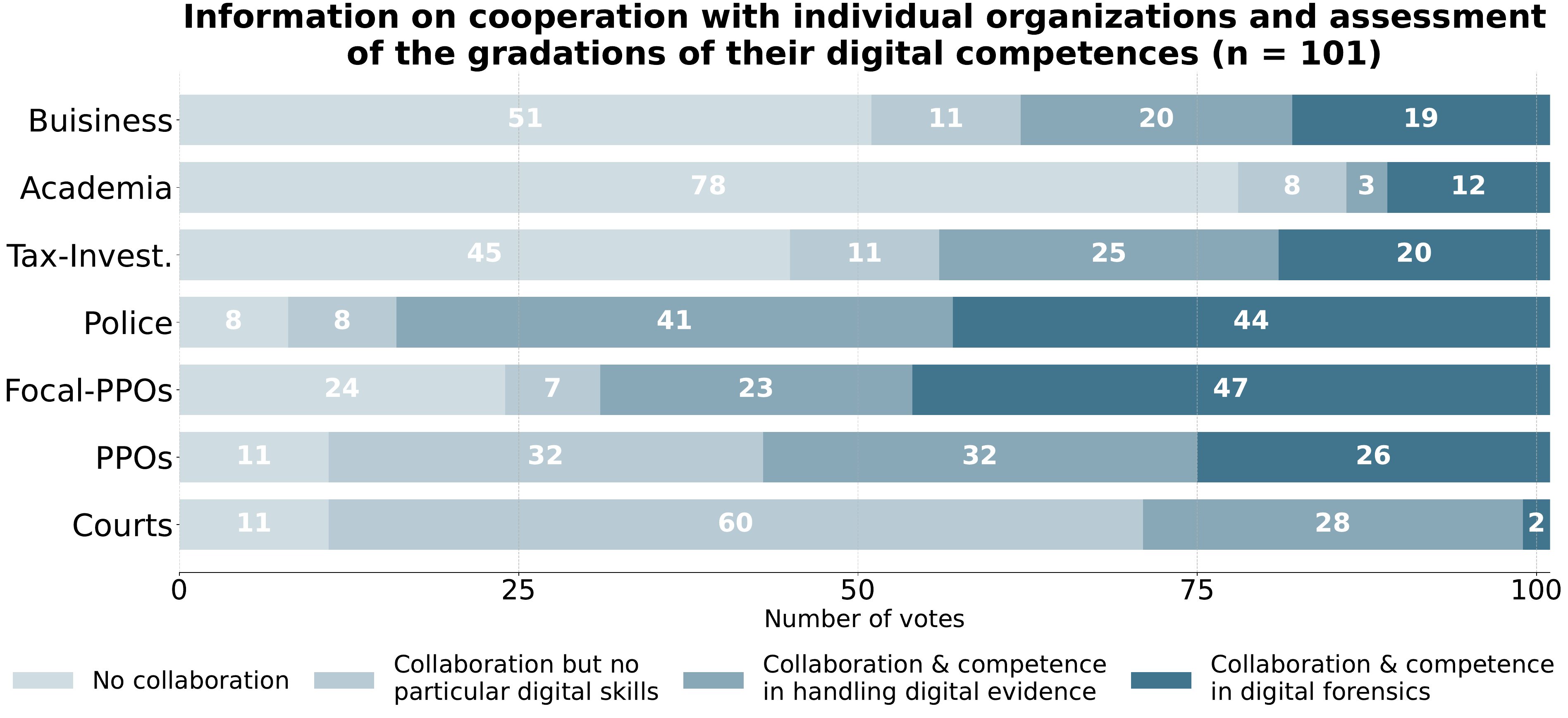}
    \Description{}
    \caption{Graphical representation of answers on cooperation and the level of expertise regarding digital evidence in various organizations involved in criminal proceedings.}
    \Description{Horizontal stacked bar chart summarizing the responses of 101 participants regarding collaboration with various organizations and their perceived level of digital forensic competence. The organizations include businesses, academic institutions, tax investigation authorities, the police, focal public prosecutors’ offices, public prosecutors’ offices, and courts. Each bar is divided into four categories: no collaboration, collaboration without specific digital expertise, collaboration with expertise in handling digital evidence, and collaboration with expertise in digital forensics. Collaboration with courts, public prosecutors’ offices, and the police is reported most frequently. However, in the case of the first two, it is predominantly associated with limited digital forensic expertise. The police, focal public prosecutors’ offices, and companies show relatively high proportions of collaboration involving expertise in handling digital evidence or in digital forensics.
}
    \label{fig: IMF_hStackedBars}
\end{figure*}

The first question in this group assessed participants' perceptions of the competencies in handling digital evidence of organizations they collaborate with. For organizations without prior collaboration, participants could select “no experience.” This excluded those organizations from subsequent questions. Three levels of competency were offered: ``no specific digital expertise'', ``expertise in the handling of digital evidence,'' and ''expertise in DF,'' with the expertise in DF as the highest level and encompassing expertise in handling digital evidence. Figure~\ref{fig: IMF_hStackedBars} visualizes the results. Participants most frequently reported collaboration with the police, followed by PPOs, courts, and focal PPOs. Slightly more than half indicated collaboration with tax investigation authorities and the business sector, while fewer than a quarter reported collaboration with academia.

In terms of perceived competencies, focal PPOs rank highest, followed by police and academia. Although the former two are closely integrated into criminal investigations, academia is also largely perceived as competent despite limited collaboration. However, more than a third of participants who collaborated with academia did not attribute particular digital competence, a proportion comparable to regular PPOs. Courts rank lowest, with around two thirds perceived as lacking notable digital competencies.

Following the survey, four questions examined perceptions of domain specific knowledge, effectiveness, efficiency, and satisfaction regarding participants’ own organizations and the other organizations listed above. All participants evaluated their own organizations, while the number of responses for other organizations varied according to the reported cooperation. 

\begin{figure*}
    \centering
    \includegraphics[width=1\linewidth]{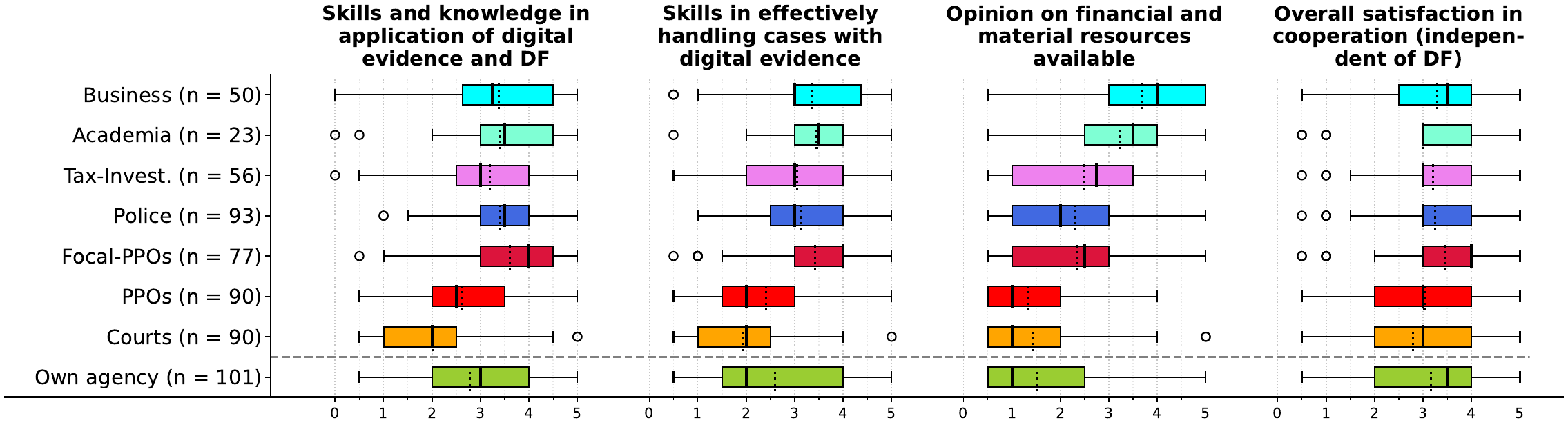}
    \Description{}
    \caption{Distribution of the results of the additional questions on competencies, case handling, resources, and overall satisfaction. The additional assessment according the respondent's perception on their own agency has been visually separated below the horizontal dotted line. The median is shown as a solid vertical line separating the middle two quartiles, and the arithmetic mean is shown as a dotted vertical line within the boxes. The variable n in the labels indicates how many participants reported collaboration and answered the questions about the respective organization.} 
    \Description{A series of box plots summarizing participants’ ratings of various organizations with respect to four aspects: skills and knowledge in the use of digital evidence and digital forensics, skills in effectively handling cases involving digital evidence, availability of financial and material resources, and overall satisfaction with collaboration, regardless of expertise in the field of digital forensics. The organizations include companies, academic institutions, tax investigation, the police, focal-PPOs, PPOs, courts, and the respondents’ own agencies. Ratings are given on a Likert scale. Each box plot shows the inter-quartile range with the median, whiskers, outliers, and the arithmetic mean. The respondents’ own agency is shown separately below a horizontal dashed line. The police, focal public prosecutors’ offices, and the respondents’ own agencies generally receive higher ratings than courts and normal public prosecutors’ offices.}
    \label{fig: DFRWS_boxplot_combi}
\end{figure*}

Figure~\ref{fig: DFRWS_boxplot_combi} reveals clear differences in the perception of competence and collaboration quality between the surveyed organizations. Focal PPOs, the police and partially tax investigation authorities receive the highest ratings regarding DF related skills, effective handling of digital evidence, and available resources.  In contrast, courts and regular PPOs are rated less positively overall, particularly with regard to effective case handling and DF competencies.

The academic sector and the private sector are also perceived comparatively positively, especially with regard to competencies and available resources, although the level of cooperation with these groups is reported to be lower. However, overall satisfaction with cooperation varies less significantly between organizations than in the other areas. Participants’ assessments of their own agencies generally fall in the middle range across all dimensions, indicating a comparatively balanced self-perception, without the pronounced positive or negative deviations observed in some external organizations.

\begin{figure*}[htbp]
    \centering
    \includegraphics[width=1\linewidth]{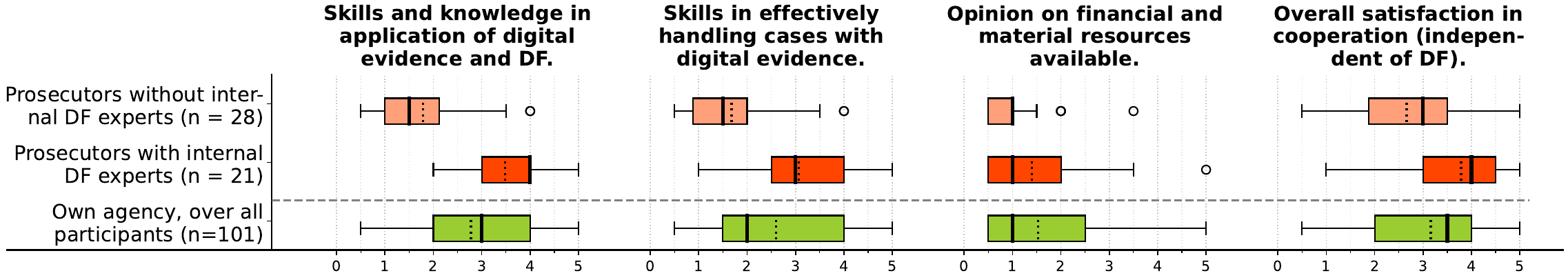}
    \Description{}
    \caption{Differentiation of the results regarding cooperation and skills within the group of public prosecutors according to their own agency's, differentiated according to whether access to own DF experts exists or not (or is unknown). Both groups include executive public prosecutors and exclude district attorneys.}
    \Description{A series of box plots comparing prosecutors’ perceptions in dependence of whether their office has access to in-house digital forensics experts or not. Three groups are presented: prosecutors without access to in-house digital forensics experts, prosecutors with access to in-house digital forensics experts, and the respondents’ overall assessment of their own office. Four aspects are rated on a Likert scale: skills and knowledge in the use of digital evidence and digital forensics, skills in effectively handling cases involving digital evidence, availability of financial and material resources, and overall satisfaction with collaboration, regardless of expertise in digital forensics. Each box plot shows the interquartile range, the median, the arithmetic mean, the whiskers, and outliers. Across all four dimensions, organizations with access to in-house digital forensics experts consistently receive higher ratings than organizations without such expertise.}
    \label{fig: DFRWS_boxplot_DFE_horizontal2}
\end{figure*}

\paragraph{\textbf{Differences between PPOs with and without DF-expertise}}
To assess the impact of internal access to DF experts, we focused on prosecutors as the sole relevant stakeholder group in the sample. Participants were filtered by professional role, access to DF expertise, and educational level, resulting in two groups: prosecutors with such access and those without (or unaware of) such access. Senior prosecutors in leadership positions (n = 8) were included and are evenly distributed in both groups. The analysis was limited to responses towards the participants’ own agency. The results are presented in Figure~\ref{fig: DFRWS_boxplot_DFE_horizontal2} alongside the overall responses.

As summarized in Table~\ref{tab: DF-Sig}, significant differences with large effect sizes in skills and knowledge, the use of digital evidence, and overall satisfaction were found, while no significant differences were observed in terms of resources. These results suggest that in our sample, access to DF expertise is associated with higher perceived competence and improved effectiveness in the handling of digital evidence, while the perception of available resources remains largely unaffected. 


\begin{table}[htbp]
  \centering
  \caption{Mann–Whitney U test results comparing groups of PPs with and without access to internal DF expertise; p-values are FDR-adjusted, effect sizes reported as r. Further details on the calculation can be found in Section~\ref{sub: Data Analysis}}
    \begin{tabular}{lllllr}
    \textbf{Category} &
      U &
      \multicolumn{1}{p{2.73em}}{\cellcolor[rgb]{ .949,  .949,  .949}p-val.\newline{}(raw)} &
      \multicolumn{1}{p{2.73em}}{p-val.\newline{}(FDR)} &
      \multicolumn{1}{p{2.73em}}{\cellcolor[rgb]{ .949,  .949,  .949}Effect\newline{}size r} &
      \multicolumn{1}{p{2.73em}}{Sig.\newline{}(FDR)}
      \\
    \midrule
    Skills \& Knowledge &
      517.0 &
      \cellcolor[rgb]{ .949,  .949,  .949}0.0000 &
      0.0000 &
      \cellcolor[rgb]{ .949,  .949,  .949}0.644 &
      \multicolumn{1}{l}{***}
      \\
    \rowcolor[rgb]{ .949,  .949,  .949} Handling Evidence &
      476.0 &
      \cellcolor[rgb]{ .851,  .851,  .851}0.0002 &
      0.0004 &
      \cellcolor[rgb]{ .851,  .851,  .851}0.525 &
      \multicolumn{1}{l}{***}
      \\
    Resources &
      328.5 &
      \cellcolor[rgb]{ .949,  .949,  .949}0.4626 &
      0.4626 &
      \cellcolor[rgb]{ .949,  .949,  .949}0.100 &
      \multicolumn{1}{l}{n.s.}
      \\
    \rowcolor[rgb]{ .949,  .949,  .949} Overall Satisfaction &
      453.5 &
      \cellcolor[rgb]{ .851,  .851,  .851}0.0011 &
      0.0014 &
      \cellcolor[rgb]{ .851,  .851,  .851}0.460 &
      \multicolumn{1}{l}{**}
      \\
       
    \end{tabular}%

  \label{tab: DF-Sig}%
\end{table}%

\subsection{Operational Challenges of Usability in DF}\label{sub: Issues related to usability in digital forensics}

This section examines various aspects of usability within DF in the context of criminal proceedings, as outlined in Section~\ref{quest: Operational Challenges and Usability Issues}. As introduced in Section~\ref{sub: The Role of Usability as an Overarching Concern}, the questions in this group address effectiveness, efficiency, and satisfaction in DF, which are particularly evident in previous research reviewed in Section~\ref{sub: Usability in Digital Forensics} and Appendix~\ref{sub: Related Issues in Digital Forensics}.

\paragraph{(a) Challenges of Digital Forensics}

\begin{figure*}
\centering

\begin{subfigure}{0.48\textwidth}
  \centering
  \includegraphics[width=\linewidth]{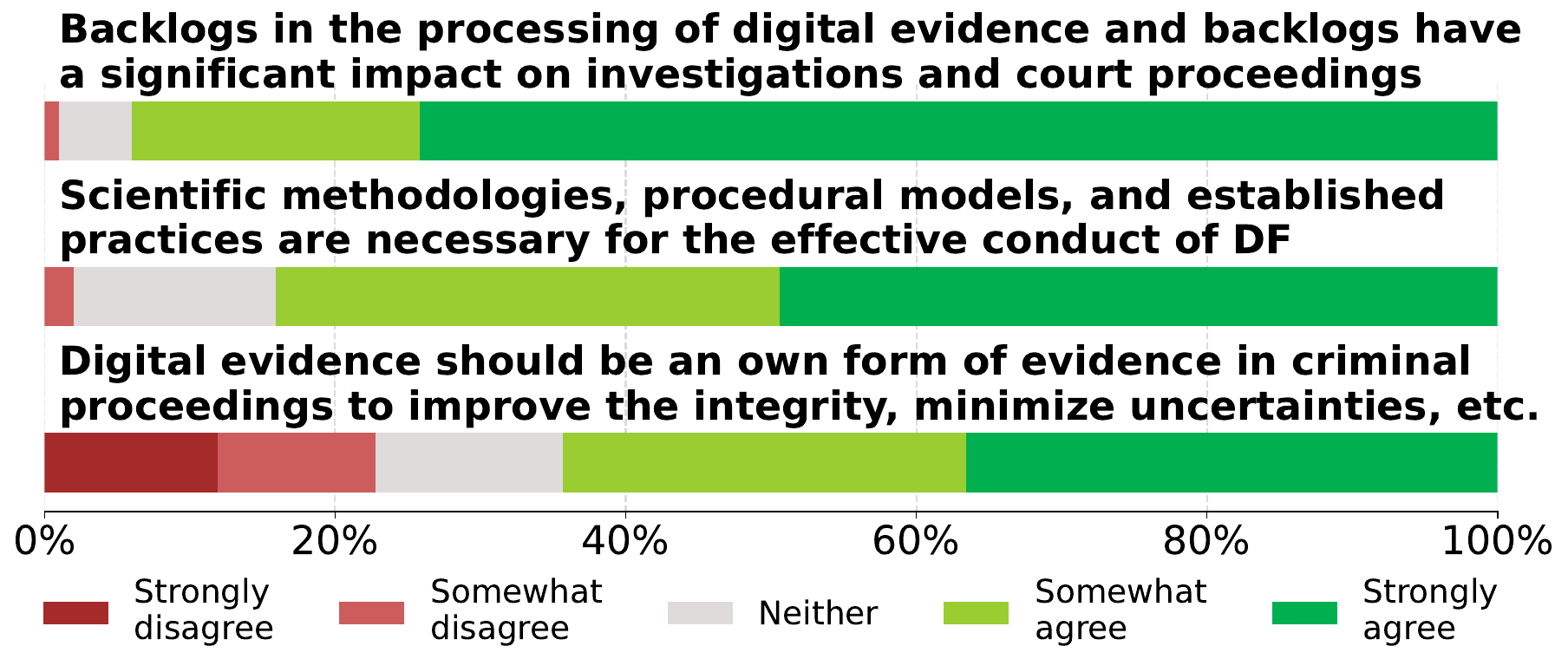}
  \vspace{1em}
  \caption{Challenges of digital forensics in criminal proceedings (n=101)\newline{}}
  \label{fig:cha}
\end{subfigure}
\hfill
\begin{subfigure}{0.48\textwidth}
  \centering
  \includegraphics[width=\linewidth]{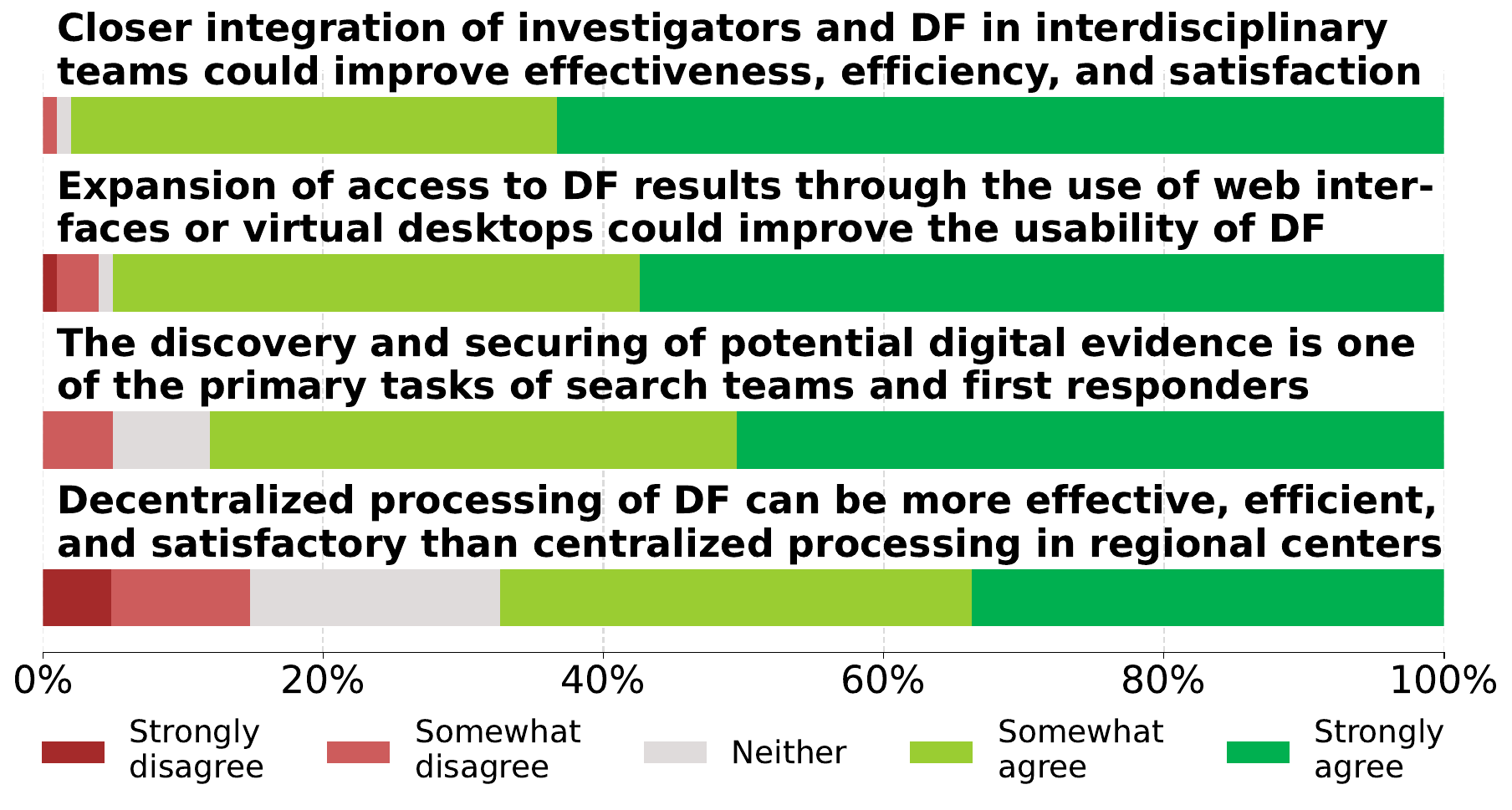}
  \caption{Collaboration in digital investigations and criminal proceedings (n=101)}
  \label{fig:col}
\end{subfigure}

\vspace{0.8em}

\begin{subfigure}{0.48\textwidth}
  \centering
  \includegraphics[width=\linewidth]{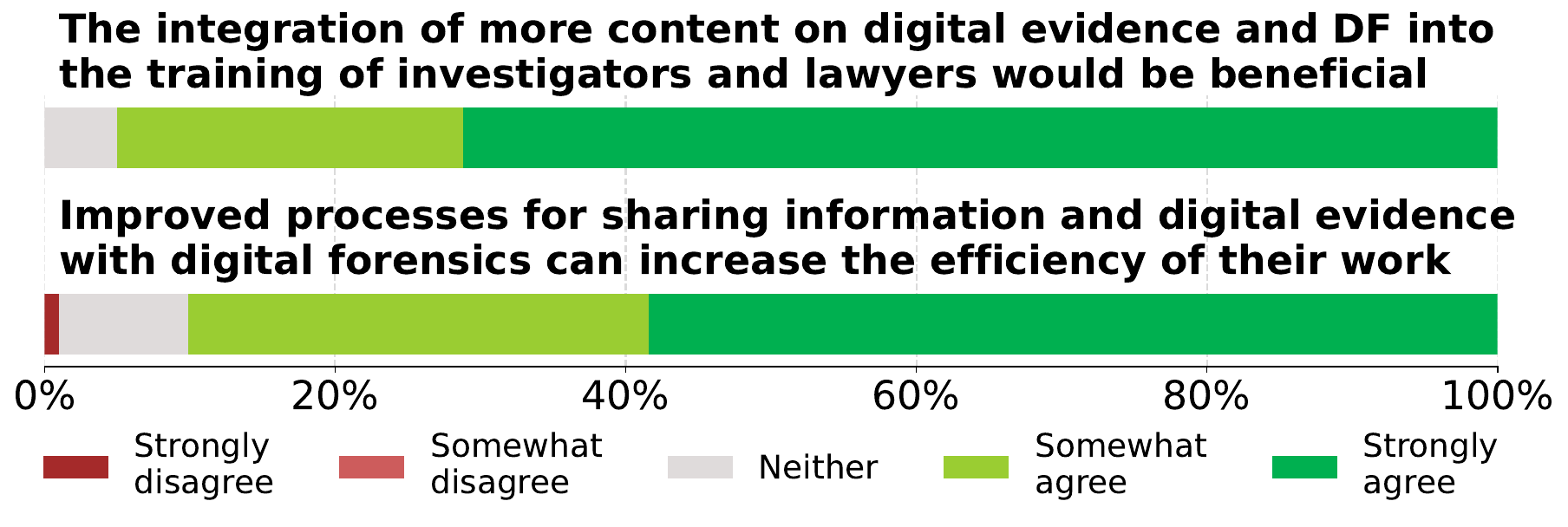}
  \caption{Information exchange and integration of DF into education (n=101)}
  \label{fig:ise}
\end{subfigure}
\hfill
\begin{subfigure}{0.48\textwidth}
  \centering
  \includegraphics[width=\linewidth]{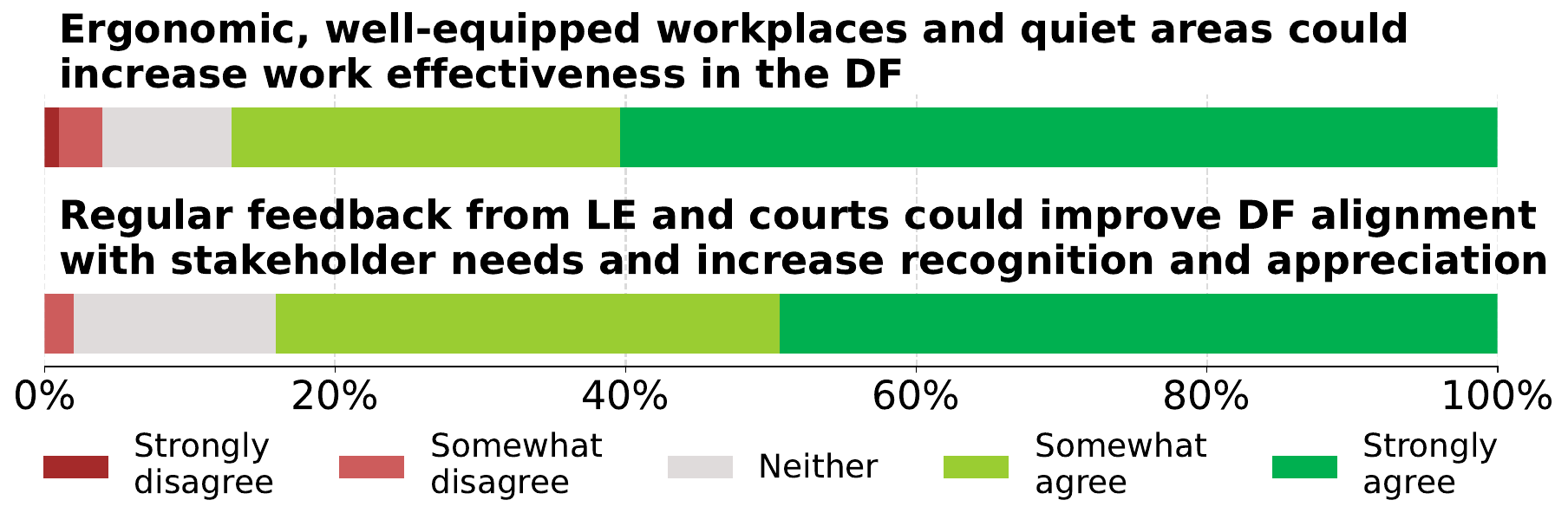}
  \caption{Satisfaction and environmental well-being (n=101)\newline{}}
  \label{fig:swb}
\end{subfigure}

\vspace{0.8em}

\begin{subfigure}{0.48\textwidth}
  \centering
  \includegraphics[width=\linewidth]{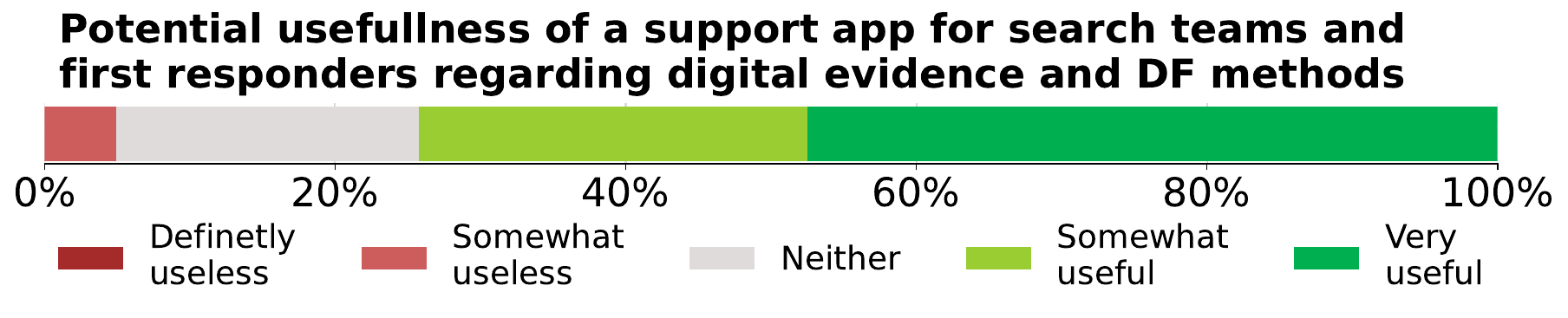}
  \caption{Support for first responders via a mobile digital evidence app (n=101)}
  \label{fig:fr}
\end{subfigure}
\hfill
\begin{subfigure}{0.48\textwidth}
  \centering
  \includegraphics[width=\linewidth]{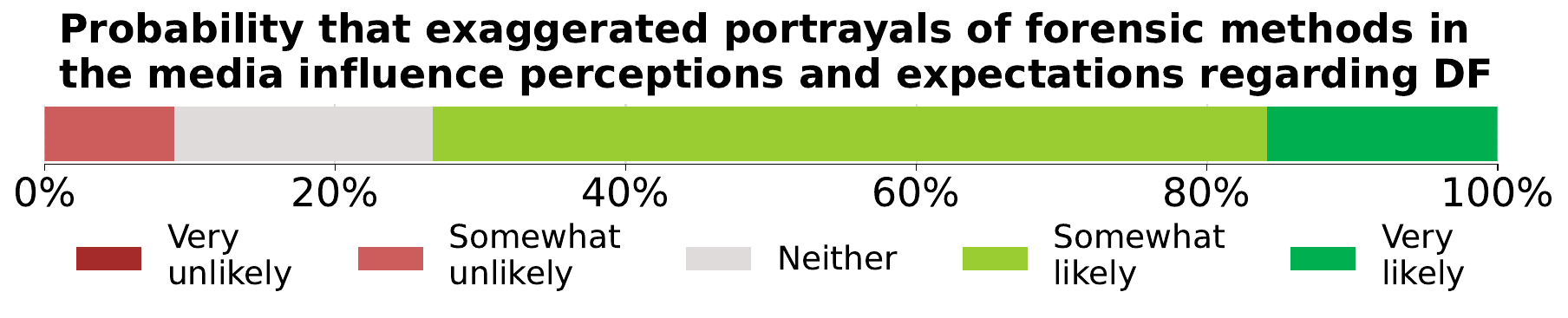}
  \caption{CSI effect in relation to DF in criminal proceedings (n=101)\newline{}}
  \label{fig:csi}
\end{subfigure}

\vspace{0.8em}

\begin{subfigure}{0.48\textwidth}
  \centering
  \includegraphics[width=\linewidth]{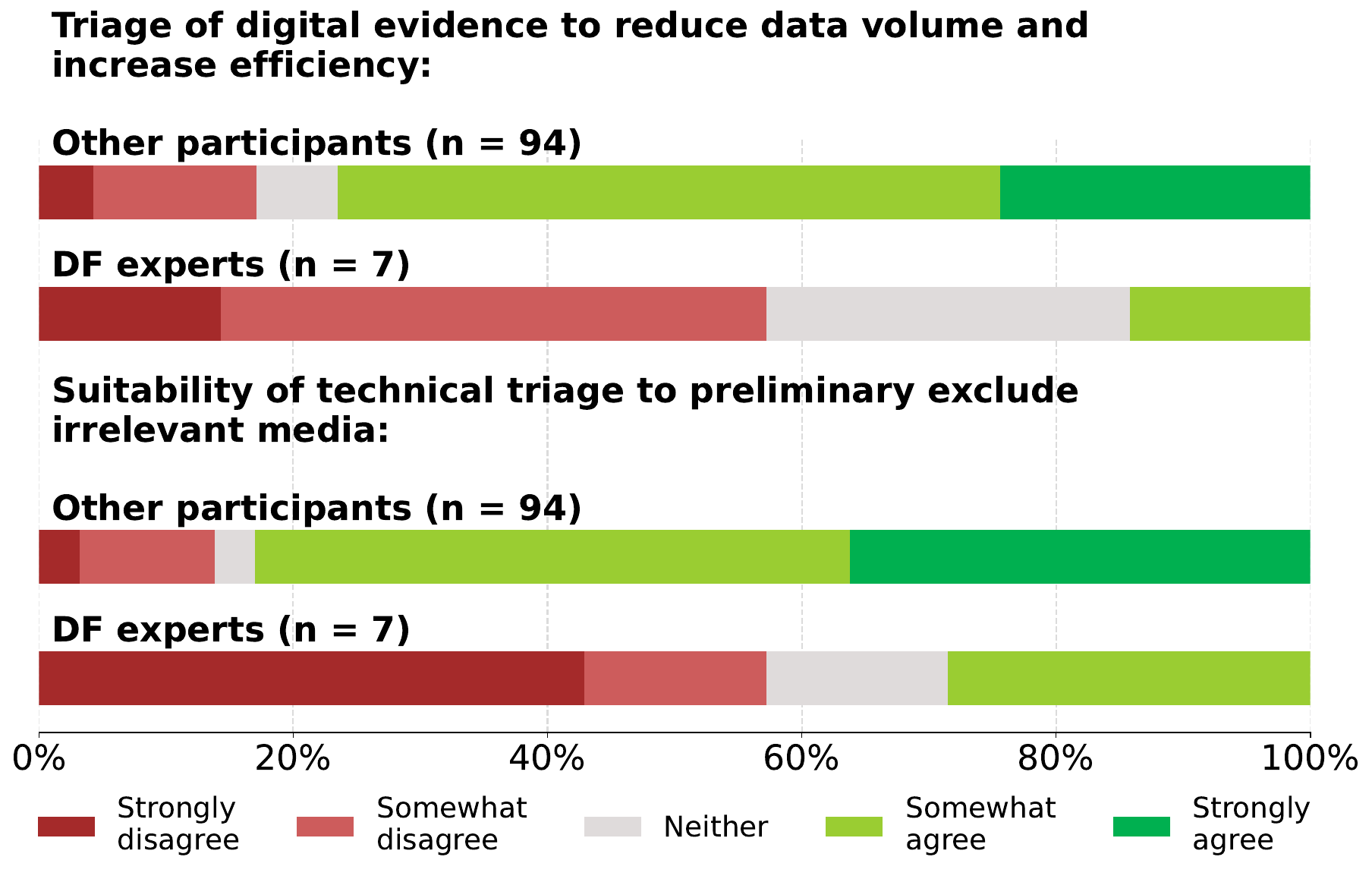}
  \caption{Views on general and technical triage (n=101)}
  \label{fig:triage}
\end{subfigure}

\caption{Results on usability-related issues in digital forensics.}
\label{fig:likert-results}
\Description{The figure consists of seven charts with horizontal bar graphs stacked on top of one another, illustrating the distribution of responses to questions—using a Likert scale—regarding aspects of the usability of digital forensics in criminal proceedings. The response categories range from ``strongly disagree'' to ``strongly agree,'' with the exception of the question on the CSI effect, where the scale ranges from ``very unlikely'' to ``very likely,'' and the question on support for first responders, where the scale ranges from ``definitely useless'' to ``useful.''
Subfigure (a) summarizes the assessments of challenges in digital forensics. Subfigure (b) shows opinions on collaboration in interdisciplinary teams. Subfigure (c) shows support for integrating digital forensics into training and improving information exchange. Subfigure (d) provides an overview of opinions on working conditions and regular feedback to improve satisfaction and effectiveness. Subfigure (e) assesses the perceived usefulness of a mobile application to support search teams and first responders. Subfigure (f) presents assessments of the influence of the CSI effect on criminal proceedings. Subfigure (g) compares the opinions of digital forensics experts and other participants regarding general and technical triage.
}
\end{figure*}

Nearly all participants report that backlogs  play a significant role in both investigations and court proceedings. In addition, they confirm that established scientific methods and practices are necessary for the effective handling of digital evidence, while more than 20\% of the participants are somewhat opposed to the idea of incorporating digital evidence into court proceedings as a separate category of admissible evidence. Figure~\ref{fig:cha} shows the breakdown of the responses to the questions in this section.

\paragraph{(b) Collaboration in Digital Forensics}

Regarding the question of whether greater interdisciplinary cooperation between police investigators and DF experts can lead to greater effectiveness, efficiency, and satisfaction, 34.7\% responded ``somewhat agree'' and 63.4\% responded ``strongly agree.''

Similar, albeit somewhat weaker, levels of agreement were found on the questions of improving usability through expanded access to DF results through web interfaces or virtual desktops —as in the case of DFaaS—and the question of whether finding, securing, or seizing digital evidence is among the primary tasks of first responders and search teams. The question of whether decentralized DF units are preferable to centralized ones received less agreement, although the majority still somewhat or fully agreed. The results are visualized in Figure~\ref{fig:col}.

\paragraph{(c) Information Sharing}\label{sub: Information Sharing}

Surprisingly, 95\% of participants support the greater integration of content from digital evidence and DF into the training and education of police investigators and legal professionals. Of these, 71.3\% of the participants strongly agree with this statement - and none gave a negative rating on this matter. As also shown in Figure~\ref{fig:ise}, the majority also agree that improved processes for the exchange of information and evidence using DF can contribute to its effectiveness.

\paragraph{(d) Satisfaction and Well-Being}

As shown in Figure~\ref{fig:swb}, high levels of agreement were also recorded on the questions about the need for adequately air-conditioned, ergonomic and quiet workspaces with quiet zones for workplace effectiveness within DF, as well as on the use of feedback to DF experts to increase perceived appreciation and better align with the needs of stakeholders.

\paragraph{(e) App Support for Search Teams and First Responders}
Regarding the question of whether participants see a benefit in providing first responders and investigative teams with a mobile app to assist in the identification, collection, and forensically sound handling of digital evidence, the majority of participants agreed, as shown in Figure~\ref{fig:fr}.

\paragraph{(f) CSI-Effect in Digital Forensics}
As Figure~\ref{fig:csi} shows, 57.4\% of the respondents consider an CSI-effect in DF somewhat likely and 15.8\% even very likely. Moreover, a majority of the relevant groups of judges, prosecutors, DF experts, investigators, and executives tend to agree that such effects are likely to have an impact.

\paragraph{(g) Triage}
The data reveal notable divergences between DF experts and the remaining participants. The results of the two questions are therefore presented with both groups visualized separately in Figure~\ref{fig:triage}. For general triage, 76.6\% of the other participants tend to agree or strongly agree that it improves efficiency, while the majority of DF experts (57.2\%) disagree or strongly disagree.

For technical triage, agreement among the other participants increases to 85\%, while the general rejection among the DF experts remains unchanged. In particular, the intensity of strong agreement or disagreement increases in both groups: from 24.5\% to 36.2\% strong agreement among the other participants and from 14.3\% to 42.9\% strong disagreement among the DF experts.

\subsection{Knowledge Management}\label{sub: KMS}

A total of 90 participants agreed that there is a need for an inter-organzational KMS that enables various professional groups to exchange information, stay informed, and receive training on DF, digital evidence, and current methods and developments. These participants were also asked two additional questions on this topic.

The results in Tables~\ref{tab: KMS1} and~\ref{tab: KMS2} present the results by group breakdown with additional heat map visualization, highlighting the most relevant functions and usage aspects. Across groups, key functions include the availability of guides and how-to materials, search features to identify experts and contacts, a knowledge base for foundational information, and access to national or international platforms. Regarding usage aspects, the primary priorities are identifying contacts and experts, accessing domain-specific knowledge, and consulting guidelines and recommended practices.

\subsection{Retaining or Recruiting DF Experts}\label{sub: Retaining or Recruiting DF Experts}

All 101 participants answered the question about the difficulty of general recruitment—specifically for the judiciary—and recruitment of DF experts. In the former case, 49.5\% of the participants responded that it is ``somewhat difficult'' to recruit staff, and 34.7\% said it is ``very difficult''. 
In the case of DF experts, 30.7\% responded that recruitment is ``somewhat difficult,'' and 55.4\% that it is ``very difficult,'' while the remaining 13.9\% selected ``neither.''

In another multiple-choice question, respondents were asked to select from nine measures to retain or attract DF experts that are possible under the aforementioned collective bargaining agreement. The results are presented in Table~\ref{tab: monetary}. In general, promotions, higher starting salaries, and the possibility of becoming a civil servant were the options most frequently selected by all participants. Despite this, allowances were also highly selected by 85.7\% of the DF experts. However, this view is shared by only 48.9\% of the other participants. In contrast, 42.2\%~of this group consider a degree program followed by a commitment to be a good measure, while this represents the least popular option among DF experts at 14.3\%.

The two questions on mutual skepticism between computer scientists and organizations of law enforcement were answered by 98 participants, regardless of group affiliation, with relatively similar trends between the different groups.
Skepticism on the part of the IT community toward law enforcement is generally considered more plausible (66.3\% consider this likely or confirm it), whereas skepticism in the opposite direction is less strongly perceived (60.2\% consider this unlikely or disagree). At the same time, it should be noted that the external perspective of computer scientists who are not part of law enforcement organizations could not be included. 

\begin{table}
  \centering
  \caption{Responses regarding measures to retain or recruit DF experts, broken down by DF experts, other groups, and the overall result, and presented as a heatmap.}
    \begin{tabular}{p{25.0em}rrr}
    \rowcolor[rgb]{ .749,  .749,  .749} \textbf{Measures to retain or recruit DF experts } & \multicolumn{1}{p{3.485em}}{\textbf{DF\newline{}(n = 7)}} & \multicolumn{1}{p{3.785em}}{\textbf{Others\newline{}(n = 90)}} & \multicolumn{1}{p{3.785em}}{\textbf{Total\newline{}(n = 97)}} \\
    \midrule
    Promotion to better positions (e.g., higher pay groups) & \cellcolor[rgb]{ 1,  .267,  .149}85.7\% & \cellcolor[rgb]{ 1,  .431,  .243}76.7\% & \cellcolor[rgb]{ 1,  .42,  .235}\textbf{77.3\%} \\
    \rowcolor[rgb]{ .949,  .949,  .949} Higher base pay or higher entry-level grades & \cellcolor[rgb]{ 1,  .529,  .298}71.4\% & \cellcolor[rgb]{ 1,  .451,  .255}75.6\% & \cellcolor[rgb]{ 1,  .459,  .259}\textbf{75.3\%} \\
    Opportunity for civil servant appointment & \cellcolor[rgb]{ 1,  .529,  .298}71.4\% & \cellcolor[rgb]{ 1,  .616,  .349}66.7\% & \cellcolor[rgb]{ 1,  .612,  .345}\textbf{67.0\%} \\
    \rowcolor[rgb]{ .949,  .949,  .949} Funding for continuing education/ degree programs & \cellcolor[rgb]{ 1,  .792,  .447}57.1\% & \cellcolor[rgb]{ 1,  .82,  .463}55.6\% & \cellcolor[rgb]{ 1,  .82,  .463}\textbf{55.7\%} \\
    Allowances (e.g., skilled worker allowance etc.) & \cellcolor[rgb]{ 1,  .267,  .149}85.7\% & \cellcolor[rgb]{ .992,  .918,  .529}48.9\% & \cellcolor[rgb]{ 1,  .894,  .502}\textbf{51.5\%} \\
    \rowcolor[rgb]{ .949,  .949,  .949} Opportunities for specialized careers & \cellcolor[rgb]{ 1,  .792,  .447}57.1\% & \cellcolor[rgb]{ .945,  .898,  .584}42.2\% & \cellcolor[rgb]{ .953,  .902,  .573}\textbf{43.3\%} \\
    Degree program with a subsequent commitment & \cellcolor[rgb]{ .749,  .824,  .812}14.3\% & \cellcolor[rgb]{ .945,  .898,  .584}42.2\% & \cellcolor[rgb]{ .929,  .894,  .6}\textbf{40.2\%} \\
    \rowcolor[rgb]{ .949,  .949,  .949} Cooperation with academia and research & \cellcolor[rgb]{ .847,  .863,  .694}28.6\% & \cellcolor[rgb]{ .843,  .859,  .702}27.8\% & \cellcolor[rgb]{ .843,  .859,  .702}\textbf{27.8\%} \\
    Advance granting of experience levels & \cellcolor[rgb]{ .949,  .902,  .576}42.9\% & \cellcolor[rgb]{ .835,  .859,  .71}26.7\% & \cellcolor[rgb]{ .843,  .859,  .702}\textbf{27.8\%} \\
      &   &   &      \\
  \textbf{Legend} (warm to cold): & \cellcolor[rgb]{ 1,  0,  0}100\% & \cellcolor[rgb]{ 1,  .922,  .518}50.0\%  & \cellcolor[rgb]{ .651,  .788,  .925}0.0\%\\
    \end{tabular}%
  \label{tab: monetary}%
\end{table}%

\begin{table*}
  \centering
  \caption{Responses of 90 participants on importance of functions of digital forensics knowledge management systems in criminal proceedings, ranked by overall agreement and shown as a heat map. The individual columns show the relative proportions by group, while the column headers also indicate the number of respondents per group.}
    \begin{tabular}{p{11em}rrrrrrrr}
    \rowcolor[rgb]{ .651,  .651,  .651} \multicolumn{1}{r}{} & \multicolumn{1}{p{3.785em}}{Judges\newline{}(n = 13)} & \multicolumn{1}{p{3em}}{PPs\newline{}(n = 40)} & \multicolumn{1}{p{3.0em}}{DF\newline{}(n = 7)} & \multicolumn{1}{p{3.0em}}{Invest.\newline{}(n = 10)} & \multicolumn{1}{p{3.0em}}{Exec.\newline{}(n = 10)} & \multicolumn{1}{p{3.65em}}{Adm./ IT\newline{}(n = 7)} & \multicolumn{1}{p{3.0em}}{Misc.\newline{}(n = 3)} & \multicolumn{1}{p{3.1em}}{\textbf{Total\newline{}(n = 90)}} \\
    \midrule
    Guides and How-Tos\newline{} & \cellcolor[rgb]{ 1,  .286,  .161}84.6\% & \cellcolor[rgb]{ 1,  .141,  .078}92.5\% & \cellcolor[rgb]{ 1,  0,  0}100.0\% & \cellcolor[rgb]{ 1,  .184,  .106}90.0\% & \cellcolor[rgb]{ 1,  .741,  .416}60.0\% & \cellcolor[rgb]{ 1,  .267,  .149}85.7\% & \cellcolor[rgb]{ 1,  .616,  .349}66.7\% & \cellcolor[rgb]{ 1,  .247,  .141}\textbf{86.7\%} \\
    \rowcolor[rgb]{ .949,  .949,  .949} Search function for experts \newline{}and contacts & \cellcolor[rgb]{ 1,  .145,  .082}92.3\% & \cellcolor[rgb]{ 1,  .278,  .157}85.0\% & \cellcolor[rgb]{ 1,  0,  0}100.0\% & \cellcolor[rgb]{ 1,  .741,  .416}60.0\% & \cellcolor[rgb]{ 1,  .369,  .208}80.0\% & \cellcolor[rgb]{ 1,  .267,  .149}85.7\% & \cellcolor[rgb]{ .882,  .875,  .655}33.3\% & \cellcolor[rgb]{ 1,  .329,  .184}\textbf{82.2\%} \\
    Knowledge databases for basic information & \cellcolor[rgb]{ 1,  .145,  .082}92.3\% & \cellcolor[rgb]{ 1,  .369,  .208}80.0\% & \cellcolor[rgb]{ 1,  0,  0}100.0\% & \cellcolor[rgb]{ 1,  .741,  .416}60.0\% & \cellcolor[rgb]{ 1,  .741,  .416}60.0\% & \cellcolor[rgb]{ 1,  .529,  .298}71.4\% & \cellcolor[rgb]{ 1,  .616,  .349}66.7\% & \cellcolor[rgb]{ 1,  .412,  .231}\textbf{77.8\%} \\
    \rowcolor[rgb]{ .949,  .949,  .949} Nationwide/ international availability & \cellcolor[rgb]{ 1,  .427,  .239}76.9\% & \cellcolor[rgb]{ 1,  .325,  .184}82.5\% & \cellcolor[rgb]{ 1,  .792,  .447}57.1\% & \cellcolor[rgb]{ 1,  .557,  .314}70.0\% & \cellcolor[rgb]{ 1,  .557,  .314}70.0\% & \cellcolor[rgb]{ 1,  .267,  .149}85.7\% & \cellcolor[rgb]{ 1,  .616,  .349}66.7\% & \cellcolor[rgb]{ 1,  .431,  .243}\textbf{76.7\%} \\
    Technical knowledge database for DF experts  & \cellcolor[rgb]{ 1,  .851,  .478}53.8\% & \cellcolor[rgb]{ 1,  .922,  .518}50.0\% & \cellcolor[rgb]{ 1,  .267,  .149}85.7\% & \cellcolor[rgb]{ 1,  .922,  .518}50.0\% & \cellcolor[rgb]{ 1,  .557,  .314}70.0\% & \cellcolor[rgb]{ 1,  .267,  .149}85.7\% & \cellcolor[rgb]{ 1,  .616,  .349}66.7\% & \cellcolor[rgb]{ 1,  .761,  .427}\textbf{58.9\%} \\
    \rowcolor[rgb]{ .949,  .949,  .949} Restricted areas for teams and government agencies & \cellcolor[rgb]{ 1,  .851,  .478}53.8\% & \cellcolor[rgb]{ .945,  .898,  .58}42.5\% & \cellcolor[rgb]{ .847,  .863,  .694}28.6\% & \cellcolor[rgb]{ 1,  .922,  .518}50.0\% & \cellcolor[rgb]{ .788,  .839,  .765}20.0\% & \cellcolor[rgb]{ 1,  .529,  .298}71.4\% & \cellcolor[rgb]{ .882,  .875,  .655}33.3\% & \cellcolor[rgb]{ .953,  .902,  .573}\textbf{43.3\%} \\
    Exchange platforms for software and program code & \cellcolor[rgb]{ .757,  .827,  .8}15.4\% & \cellcolor[rgb]{ .929,  .894,  .6}40.0\% & \cellcolor[rgb]{ 1,  .529,  .298}71.4\% & \cellcolor[rgb]{ 1,  .922,  .518}50.0\% & \cellcolor[rgb]{ .929,  .894,  .6}40.0\% & \cellcolor[rgb]{ 1,  .529,  .298}71.4\% & \cellcolor[rgb]{ .882,  .875,  .655}33.3\% & \cellcolor[rgb]{ .945,  .898,  .584}\textbf{42.2\%} \\
    \rowcolor[rgb]{ .949,  .949,  .949} Involvement of the academic community & \cellcolor[rgb]{ .918,  .89,  .612}38.5\% & \cellcolor[rgb]{ .753,  .827,  .804}15.0\% & \cellcolor[rgb]{ 1,  .792,  .447}57.1\% & \cellcolor[rgb]{ .788,  .839,  .765}20.0\% & \cellcolor[rgb]{ .718,  .812,  .847}10.0\% & \cellcolor[rgb]{ .949,  .902,  .576}42.9\% & \cellcolor[rgb]{ .882,  .875,  .655}33.3\% & \cellcolor[rgb]{ .82,  .851,  .729}\textbf{24.4\%} \\
    Digital awards and rating features & \cellcolor[rgb]{ .812,  .847,  .737}23.1\% & \cellcolor[rgb]{ .753,  .827,  .804}15.0\% & \cellcolor[rgb]{ .651,  .788,  .925}0.0\% & \cellcolor[rgb]{ .788,  .839,  .765}20.0\% & \cellcolor[rgb]{ .718,  .812,  .847}10.0\% & \cellcolor[rgb]{ .847,  .863,  .694}28.6\% & \cellcolor[rgb]{ .651,  .788,  .925}0.0\% & \cellcolor[rgb]{ .757,  .827,  .8}\textbf{15.6\%} \\
    \rowcolor[rgb]{ .949,  .949,  .949} Involvement of business and industry & \cellcolor[rgb]{ .702,  .808,  .863}7.7\% & \cellcolor[rgb]{ .702,  .808,  .867}7.5\% & \cellcolor[rgb]{ .749,  .824,  .812}14.3\% & \cellcolor[rgb]{ .788,  .839,  .765}20.0\% & \cellcolor[rgb]{ .651,  .788,  .925}0.0\% & \cellcolor[rgb]{ .949,  .902,  .576}42.9\% & \cellcolor[rgb]{ .882,  .875,  .655}33.3\% & \cellcolor[rgb]{ .733,  .82,  .827}\textbf{12.2\%} \\
      &   &   &   &   &   &   &   &  \\
    & \textbf{Legend:} & warm & \cellcolor[rgb]{ 1,  0,  0}100\% & \cellcolor[rgb]{ 1,  .463,  .259}75.0\% & \cellcolor[rgb]{ 1,  .922,  .518}50.0\% & \cellcolor[rgb]{ .824,  .855,  .722}25.0\% & \cellcolor[rgb]{ .651,  .788,  .925}0.0\% & cold \\

    \end{tabular}%
  \label{tab: KMS1}%
\end{table*}%

\begin{table*}
  \centering
  \caption{Responses from 89 participants regarding which aspects of  digital forensics knowledge management systems usage would be most important to them.  The individual columns show the relative proportions by group, while the column headers also indicate the number of respondents per group}
    \begin{tabular}{p{11em}rrrrrrrr}
    \rowcolor[rgb]{ .651,  .651,  .651} \multicolumn{1}{r}{} & \multicolumn{1}{p{3.785em}}{Judges\newline{}(n = 13)} & \multicolumn{1}{p{3.0em}}{PPs\newline{}(n = 39)} & \multicolumn{1}{p{3.0em}}{DF\newline{}(n = 7)} & \multicolumn{1}{p{3.0em}}{Invest.\newline{}(n = 10)} & \multicolumn{1}{p{3.0em}}{Exec.\newline{}(n = 10)} & \multicolumn{1}{p{3.75em}}{Adm./ IT\newline{}(n = 7)} & \multicolumn{1}{p{3.0em}}{Misc.\newline{}(n = 3)} & \multicolumn{1}{p{3.1em}}{\textbf{Total\newline{}(n = 89)}} \\
    \midrule
    Finding contacts/ experts\newline{} & \cellcolor[rgb]{ 1,  .145,  .082}92.3\% & \cellcolor[rgb]{ 1,  .325,  .184}82.5\% & \cellcolor[rgb]{ 1,  .267,  .149}85.7\% & \cellcolor[rgb]{ 1,  0,  0}100.0\% & \cellcolor[rgb]{ 1,  .184,  .106}90.0\% & \cellcolor[rgb]{ 1,  .529,  .298}71.4\% & \cellcolor[rgb]{ .882,  .875,  .655}33.3\% & \cellcolor[rgb]{ 1,  .29,  .165}\textbf{84.4\%} \\
    \rowcolor[rgb]{ .949,  .949,  .949} Accessing information on specific subject areas & \cellcolor[rgb]{ 1,  .286,  .161}84.6\% & \cellcolor[rgb]{ 1,  .647,  .365}65.0\% & \cellcolor[rgb]{ 1,  0,  0}100.0\% & \cellcolor[rgb]{ 1,  .922,  .518}50.0\% & \cellcolor[rgb]{ 1,  .557,  .314}70.0\% & \cellcolor[rgb]{ 1,  .267,  .149}85.7\% & \cellcolor[rgb]{ 1,  0,  0}100.0\% & \cellcolor[rgb]{ 1,  .514,  .29}\textbf{72.2\%} \\
    Accessing guidelines and recommendations & \cellcolor[rgb]{ 1,  .71,  .4}61.5\% & \cellcolor[rgb]{ 1,  .416,  .235}77.5\% & \cellcolor[rgb]{ 1,  .529,  .298}71.4\% & \cellcolor[rgb]{ 1,  .184,  .106}90.0\% & \cellcolor[rgb]{ 1,  .369,  .208}80.0\% & \cellcolor[rgb]{ .949,  .902,  .576}42.9\% & \cellcolor[rgb]{ .651,  .788,  .925}0.0\% & \cellcolor[rgb]{ 1,  .533,  .302}\textbf{71.1\%} \\
    \rowcolor[rgb]{ .949,  .949,  .949} Opportunities for self-development & \cellcolor[rgb]{ 1,  .427,  .239}76.9\% & \cellcolor[rgb]{ 1,  .741,  .416}60.0\% & \cellcolor[rgb]{ 1,  .792,  .447}57.1\% & \cellcolor[rgb]{ 1,  .557,  .314}70.0\% & \cellcolor[rgb]{ .859,  .867,  .682}30.0\% & \cellcolor[rgb]{ .749,  .824,  .812}14.3\% & \cellcolor[rgb]{ 1,  .616,  .349}66.7\% & \cellcolor[rgb]{ 1,  .8,  .451}\textbf{56.7\%} \\
    Staying informed about new developments & \cellcolor[rgb]{ .918,  .89,  .612}38.5\% & \cellcolor[rgb]{ 1,  .878,  .494}52.5\% & \cellcolor[rgb]{ 1,  .792,  .447}57.1\% & \cellcolor[rgb]{ 1,  .557,  .314}70.0\% & \cellcolor[rgb]{ 1,  .741,  .416}60.0\% & \cellcolor[rgb]{ 1,  .792,  .447}57.1\% & \cellcolor[rgb]{ 1,  0,  0}100.0\% & \cellcolor[rgb]{ 1,  .82,  .463}\textbf{55.6\%} \\
    \rowcolor[rgb]{ .949,  .949,  .949} Exchange with people from different areas of expertise & \cellcolor[rgb]{ 1,  .71,  .4}61.5\% & \cellcolor[rgb]{ .894,  .878,  .643}35.0\% & \cellcolor[rgb]{ 1,  .529,  .298}71.4\% & \cellcolor[rgb]{ .929,  .894,  .6}40.0\% & \cellcolor[rgb]{ 1,  .922,  .518}50.0\% & \cellcolor[rgb]{ .949,  .902,  .576}42.9\% & \cellcolor[rgb]{ .651,  .788,  .925}0.0\% & \cellcolor[rgb]{ .953,  .902,  .573}\textbf{43.3\%} \\
    Exchange with people from your own area of expertise & \cellcolor[rgb]{ .918,  .89,  .612}38.5\% & \cellcolor[rgb]{ .808,  .847,  .745}22.5\% & \cellcolor[rgb]{ 1,  .792,  .447}57.1\% & \cellcolor[rgb]{ .929,  .894,  .6}40.0\% & \cellcolor[rgb]{ 1,  .922,  .518}50.0\% & \cellcolor[rgb]{ .949,  .902,  .576}42.9\% & \cellcolor[rgb]{ .651,  .788,  .925}0.0\% & \cellcolor[rgb]{ .882,  .875,  .655}\textbf{33.3\%} \\
      &   &   &   &   &   &   &   &  \\
    & \textbf{Legend:} & warm & \cellcolor[rgb]{ 1,  0,  0}100\% & \cellcolor[rgb]{ 1,  .463,  .259}75.0\% & \cellcolor[rgb]{ 1,  .922,  .518}50.0\% & \cellcolor[rgb]{ .824,  .855,  .722}25.0\% & \cellcolor[rgb]{ .651,  .788,  .925}0.0\% & cold \\
    \end{tabular}%
  \label{tab: KMS2}%
\end{table*}%
\subsection{Concluding Question}\label{sub: Concluding Question}

In conclusion, 93.1\% of participants stated in a final question that usability in DF within criminal proceedings will play an increasing role in the future. 
In addition, all 101 participants whose responses were included in the results confirmed that they had completed the survey in good faith and that their responses could be included in the study.

\section{Discussion}
\label{sec:discussion}

In the following, we discuss the results of the survey, research questions, limitations, and future work.

\subsection{Discussion of the Survey Results}\label{sub: Discussion of the Survey Results}

\subsubsection{Demographics}\label{dis: Demographics}
The sample reflects several structural characteristics of the German judicial system that are relevant to interpret the results. Positions involving sovereign authority, such as judges and prosecutors, are typically held by civil servants, whereas DF experts, investigators, and administrative or IT personnel are often employed. 
Strong representation of participants with advanced academic qualifications is not surprising given the educational requirements of the German judiciary, where most professional careers require qualifications comparable to a master’s degree. In the legal domain, this includes the German State Examinations, with the Second State Examination constituting the formal prerequisite for judge and prosecutor careers.

\subsubsection{Training and Access to Expertise}\label{dis: Training and Access to Expertise}

The reported participation rate in DF-related training in the past five years~(45\%) initially appears to be lower than the 56\% reported in Flory’s decade-old study~\citep{Flory.2016}. However, the latter measured whether agencies had sent at least one employee to training, whereas the present study captures individual participation. Consequently, actual participation at the individual level is likely higher in the sample, although direct comparisons remain limited.

The training content related to DF is dominated by the three technical domains (see Table~\ref{tab: training}). Excluding technical user groups such as DF experts and administration/IT, only criminalistic aspects enter the top three, while technical cybercrime/IT security drops out. This focus on more technical topics is somewhat unexpected, given that only a small proportion of participants have a technical background. However, it should also be noted that the most frequently aforementioned topic, digital evidence, encompasses both technical and legal dimensions, although it was classified as technical due to the nature of evidence.

Furthermore, more than a third of the participants who are not experts in DF themselves confirmed that they have internal access to these experts (see Table~\ref{tab: training}). Given the limited prevalence of such units within the judiciary compared to the police, this seems surprising. At the same time, we assume that participants with ties to DF may have been more likely to participate in the survey than those without it.

\subsubsection{Digital Skills, Capabilities, and Competencies} \label{dis: Digital Skills, Capabilities, and Competencies}

The results of Section~\ref{sub: Collaboration and Digital Skills} and Figure~\ref{fig: DFRWS_boxplot_combi} indicate that the collaboration is strongest among courts, PPOs, focal PPOs, and police, with high DF competence primarily attributed to the latter two. Courts rank notably lower in this regard, which may be explained by their lack of regular own investigative activities. In contrast, collaboration with academia is comparatively limited, while perceptions of its digital competence are polarized—ranging from lack of particular expertise to full DF competence. This divergence may reflect the broad and heterogeneous expertise within the academic domain.

Regarding expectations of skills and knowledge, effective case handling, and available resources, Fig.~\ref{fig: IMF_hStackedBars} indicates consistently stronger performance for organizations typically associated with DF expertise (focal PPOs, police, tax investigators) and for actors outside traditional criminal proceedings (academia, private sector). In contrast, overall satisfaction remains relatively uniform, supporting its relative independence from the other usability criteria, as described by~\citet{ISO.2018}.

A more differentiated picture emerges for resource assessments when considering prosecutors with internal DF access (Fig.~\ref{fig: DFRWS_boxplot_combi}), who are likely the best positioned to evaluate focal PPOs. Here, ratings align more closely with those of regular PPOs, suggesting that perceptions of external organizations may be comparatively more favorable when not based on direct experience.

\subsubsection{Effects of Own DF Experts on PPOs} \label{dis: Effects of Own DF Experts on PPOs}

The effects described in Section~\ref{sub: Collaboration and Digital Skills} and illustrated in Figure~\ref{fig: DFRWS_boxplot_DFE_horizontal2} and Table~\ref{tab: DF-Sig} indicate that PPOs with internal DF units exhibit consistently more favorable outcomes than those without such internal expertise. With the exception of aspects related to resource allocation, access to internal DF expertise is associated with clearly positive effects. These differences are particularly pronounced and statistically significant in the domains of perceived competencies and effectiveness in handling digital evidence, while a somewhat weaker, yet still observable, level of significance is found with regard to overall satisfaction in cooperation.

 The relatively small number of responses and the imbalance in group sizes limit the generalizability of these results. Nevertheless, as such contrasts have become increasingly rare due to the widespread establishment of DF units—particularly within police organizations—the observed effects are noteworthy, and provide valuable insight into the impact of internal expertise structures on the competencies in handling digital evidence and DF. 

\subsubsection{Perception of Challenges} \label{dis: Perception of Challenges}

Strong confirmation of the problem of  persistent backlogs in DF is not surprising given their impact on downstream judicial processes, particularly when deadlines in criminal proceedings cannot be met~\citep{Pollitt.2013,WilsonKovacs.2020,Horsman.2022}. At the same time, the high level of agreement on the need for scientific methods, process models, and established practices highlights their importance for the effective application of DF in criminal proceedings~\citep{Palmer.2001,Casey.2011,Dewald.2015}. This may indicate a growing recognition of DF as a scientific discipline within the judiciary and an acceptance of structured scientific approaches.

The introduction of digital evidence as a distinct category in court proceedings~\citep{Gless.2021} is considered to be more controversial. Judges, in particular, show higher levels of opposition, with 26.7\% strongly and 13.3\% somewhat disagreeing, although the majority still express support. This pattern is not surprising, as such a change would primarily affect court proceedings, while it could potentially simplify the work of investigators and prosecutors who must adapt digital evidence to fit strict evidence categories (see Section~\ref{sub: Specialties in German Criminal Proceedings}).

\subsubsection{Collaboration in Digital Forensics} \label{dis: Collaboration in Digital Forensics}

The results of Section~\ref{sub: Issues related to usability in digital forensics} and Figure~\ref{fig:col} show a clear preference among participants for greater interdisciplinary collaboration between traditional investigators and DF experts~\citep{Hansen.2017}, while at the same time there is support for expanding access to DF results, e.g., via DFaaS solutions~\citep{Beek.2020}. Furthermore, the strong confirmation that searching and securing of digital evidence is one of the primary tasks of first responders and search teams~\citep{Bossler.2012,WilsonKovacs.2020,Warner.2026} indicates that this cooperation should not only be viewed as a one-way street from DF to other groups involved in criminal proceedings. Finally, the preference for decentralized DF units could indicate that local and direct points of contact are preferred over a central, e.g., state-wide, unit for this cooperation~\citep{Casey.2019}.

\subsubsection{Education and Information exchange} \label{dis: Education and Information exchange}

As previously described, participation rates in DF related training remain comparatively low, particularly among judges and PPs. Against this background, the strong participant's support for the integration of digital evidence and DF in the education and continuing training of legal professionals and investigators, visualized in Figure~\ref{fig:ise} is noteworthy~\citep{Flory.2016, Henseler.2018}. This may indicate a perceived urgent need for action resulting from the increasing prevalence of digital evidence in criminal proceedings~\cite{EU.2025}.

Development of corresponding educational concepts could particularly benefit from usability and human-centered security research. Similarly to IT security awareness approaches, as described, e.g., by~\citet{Sasse.2023c}, effective training must go beyond the simple transfer of knowledge, and develop standardized routines for the forensically sound handling of digital evidence and DF, and behavioral conduct of the actors involved. Parallels to IT security awareness can also be observed in the susceptibility to biases discussed in Section~\ref{sub: Examples of Satisfaction and UX in DF}, which could potentially be mitigated through the development of DF awareness. Another important aspect is the development of self efficacy according to DF, which plays a significant role for first responders, DF experts, decision makers, and stakeholders throughout DF in criminal proceedings.

Perceived importance of improving information exchange with respect to digital evidence and DF~\citep{Nouh.2019} could also benefit from this integration into education and expanded training, as this can foster stronger shared understanding between stakeholder groups, including DF.

\subsubsection{Satisfaction and Well-Being} \label{dis: Satisfaction and Well-Being}

Participants also strongly confirm the importance of environmental conditions on the job for the effectiveness of DF work~\citep{Kelty.2021}. More than 60\% strongly agree that factors such as ergonomics, equipment, climate control, and quiet work environments can positively impact effectiveness. This may indicate a growing awareness of the technical, physical, and mental demands associated with DF. 

Feedback emerges as an important factor in relation to collaboration~\citep{Strickland.2023}. Its perceived effectiveness in improving the alignment between the DF units and the requirements of stakeholders, as well as in enhancing the recognition of the DF experts, represents a positive finding, especially in the sense of collaboration. However, more research is needed to determine how such feedback mechanisms can be implemented in criminal proceedings. It is also important to note that feedback may not always have positive effects. Negative feedback or unfavorable case outcomes may lead to frustration. Consequently, design of such systems is complex and requires careful and responsible implementation.

\subsubsection{Triage} \label{dis: Triage}

Triage and technical triage seem to hit a nerve on the organizational side of DF. As shown in Figure~\ref{fig: DFRWS_boxplot_combi} or in the literature, there is often a lack of resources in DF and investigations~\citep{Pollitt.2010, Nouh.2019, Warner.2026, WilsonKovacs.2020}. Therefore, reducing the efforts seems to be a good way to deal with this. However, especially technical triage comes with its own efforts, as shown by~\citet{WilsonKovacs.2020}. If practiced at the search location, it can cause additional stress. 

If a small number of police officers are responsible not only for technical triage but  simultaneously, for searches, interrogations, and securing the scene, this can quickly exhaust their capacity and push them to their limits. Stress can also arise in such situations even when the work is carried out by DF experts, for example, if instructions for carrying out the work conflict with their own experience and observations or if additional time is needed for preliminary analyzes, in which case, when conducted by private-sector DF experts, additional security measures must be provided by police officers. This can be compounded by doubts about the effectiveness of technical triage or, as described by~\citet{Warner.2026}, the circumvention of such measures through workarounds.

According to the results shown in Figure~\ref{fig:triage}, it should also be noted that DF experts in the sample are likely the only group directly affected by the implementation of such measures. Unlike other investigators in the sample, who most probably have expertise in the field of economics, DF experts must, in cases of doubt, either apply the technology themselves or accept the results from other groups without being able to influence them. However, as~\citet{WilsonKovacs.2020} shows, the responsibility is often attributed to DF experts in both cases. Therefore, the greater skepticism among DF experts seems understandable, while for the non-involved groups, the idea of reducing the workload or the amount of material to be analyzed may be the primary concern.

\subsubsection{Support for First Responders and the CSI Effect} \label{dis: Support for First Responders and the CSI Effect}

As described in Sections~\ref{sub: Usability in Digital Forensics} and~\ref{sub: Examples of Satisfaction and UX in DF}, securing or seizing of digital evidence, often carried out by non-technical first responders, represents a critical point at the beginning of the chain-of-custody and DF processes~\citep{Bossler.2012,WilsonKovacs.2020}. The benefits of a support app on mobile devices for these groups, as suggested by~\citet{Casey.2021}, were confirmed by the large majority of the participants. This also appears to support the findings of~\citet{Warner.2026} and~\citet{WilsonKovacs.2020} that existing training and assistance measures are insufficient and that more practical support must be developed.

Responses shown in Figure~\ref{fig:csi} appear to support the presence of a CSI effect in DF, which can also be a source of bias~\citep{Cole2009,Sunde.2019,Sunde.2021}. However, these findings should be interpreted with caution: perceiving such an influence as ``somewhat likely'' or ``very likely’’ does not necessarily imply its actual occurrence. Rather, the results may also reflect increased awareness of the phenomenon, which has been discussed by lawyers for more than two decades. Although more data is needed to assess the actual prevalence more conclusively, the results also reinforce the importance of education~\citep{Hansen.2017,Henseler.2018} and awareness of key stakeholders in the field of DF. For example, DF education can help raise awareness of such effects and the actual capabilities of DF.

\subsubsection{Knowledge Management} \label{dis: Knowledge Management}

In all groups, a consistently high demand for DF-KMS is observed~\citep{Casey.2021}. Core functions such as procedural guidelines, expert search functions, and knowledge databases appear to be universally relevant, suggesting shared needs between all stakeholders. At the same time, the results point to differentiated requirements that reflect role-specific priority areas, for example, a stronger emphasis on interdisciplinary exchange in groups of judges and DF experts, as well as a greater demand for software-based exchange platforms among DF experts and the administration/IT (see Tables~\ref{tab: KMS1} and~\ref{tab: KMS2}). The results offer new insights that may be useful in adapting such systems for a wider user group in criminal proceedings.

\subsubsection{Retaining or Recruiting DF Experts} \label{dis: Retaining or Recruiting DF Experts}

Although questions regarding the retention or recruitment of DF personnel are largely tied to the national-specific and predominantly monetary provisions of collective bargaining agreements, they can contribute to current discussions on well-being in the field of DF research. In addition to better pay, factors such as recognition and appreciation can also play a role, for example, in civil servant status, promotions, higher positions, or specialized careers~\citep{Kelty.2021,Baumgartner.2023,Strickland.2023}. Furthermore, nearly all participants answered the question, which also provides insight into the organizational perspective and reveals differences, such as the rating of the allowances (see Table~\ref{tab: monetary}).

Findings on the difficulty of recruiting new staff indicate that significant challenges are perceived in recruiting both general legal staff and DF experts, although recruiting DF experts is perceived to be slightly more challenging. Both areas face competition from other external organizations, which often offer higher financial incentives and can only select from a limited pool of applicants.

Our questions on the relationship between law enforcement and computer scientists was based on questions tailored to the German legal system. However, these results should be interpreted with caution, as our sample reflects only the perspective of the judiciary (see Section~\ref{sub: Retaining or Recruiting DF Experts}).

\subsubsection{The Future Role of Usability in Digital Forensics} \label{dis: The Future Role of Usability in Digital Forensics}

We consider that the integration of a brief explanation of usability, along with the survey’s extensive and wide-ranging topics, enabled participants to provide a sound assessment of whether usability in DF will play a greater role in criminal proceedings in the future. The high levels of agreement on this question indicate that the topic of usability is viewed as relevant in the fields of DF, digital investigations, and digital evidence (see Section~\ref{sub: Concluding Question}).

\subsection{Discussion of the Research Questions}\label{sub: Research Questions Revisited}
\paragraph{RQ1: Perceptions of Digital Forensics by Different Stakeholders} 
 
The survey participants, who were divided into different groups based on their roles as shown in Table~\ref{tab: demo}, all originate from the judiciary.
Other authorities relevant to criminal proceedings, such as the police or tax investigators, are not included in the sample. 
This could be one reason why the differences in  responses to the questions are often minor.

Still, several interesting differences emerged.
Most notably, there are differences in the assessment of ``cooperation and skills’’ between PPs with and without access to internal DF expertise (see Section~\ref{dis: Effects of Own DF Experts on PPOs}), while DF experts and the remaining participants  differed in their perceptions of the efficiency and effectiveness of different triage measures (see Section~\ref{dis: Triage}). Minor differences were observed regarding the recruitment and retention of DF experts and the perceived need for KMSs. Participants also showed preferences, e.g., for decentralized DF departments or preferred functions of KMSs (see Sections~\ref{dis: Knowledge Management} and~\ref{dis: Retaining or Recruiting DF Experts}).

\paragraph{RQ2: Perceived Relevance of Usability Challenges in Digital Forensics by Stakeholders} The results also confirm the relevance of the identified usability challenges or reveal clear trends regarding preferred solutions or priorities. Participants confirmed, e.g., DF-backlogs as a major challenge in criminal proceedings and confirmed the need for KMSs (see Sections~\ref{dis: Perception of Challenges} and~\ref{dis: Knowledge Management}). This is also reflected in the confirming results of our concluding question regarding the future relevance of usability in the DF (see Section~\ref{dis: The Future Role of Usability in Digital Forensics}). Overall, the results indicate that the identified challenges are relevant across stakeholder groups.

\paragraph{The Interconnection Between the Research Questions and Usability} 
As described in Section~\ref{sub: The Role of Usability as an Overarching Concern}, most of the challenges addressed in the survey questions were identified in literature and assigned to one or more of the three usability criteria: effectiveness, efficiency, and satisfaction. 
Together with the participants’ reported perceptions, the survey provides a broad picture of the challenges related to the usability of DF in criminal proceedings from the perspective of users in the German judicial system. 

\subsection{Limitations}

This study has the following limitations.

\paragraph{Sampling Bias and Coverage Limitations}
A potential limitation is self-selection bias, which may have led to an over-representation of participants with a strong interest in the topic. The findings may also be affected by regional sampling bias, as the data was collected exclusively from authorities within a single German federal state. Given structural and organizational differences, the results may not fully generalize to other regions.

Although the study was originally intended to include a wider range of law enforcement agencies, the participation of additional agencies could not be achieved. Nevertheless, the data collected provide valuable and in-depth insights into the perspectives of judicial authorities, which are often underrepresented in DF research.

\paragraph{Survey Design}

The survey design may introduce acceptance bias, as several elements were formulated as agreement statements (e.g. ``Do you agree that…’'). This is particularly relevant for the questions described in Section~\ref{sub: Issues related to usability in digital forensics}. However, this was a conscious decision derived from design considerations. Since the preceding question set is cognitively demanding and the items can introduce unfamiliar or novel aspects of DF or re-frame existing ones in a new context, participants were encouraged to provide their subjective opinion rather than a rating. Although this approach aimed to accommodate the survey's complexity, the high agreement rates across nearly all items suggest that the potential influence of leading questions must be considered.

\paragraph{Data Errors}

A data entry caught our attention during the review. One participant reported being a judge and also being employed. These are two mutually exclusive conditions, since judges are always civil servants. Since this was only one of two entries submitted through QR code access and all other data were valid, the data were retained. We assume that the error may have been caused by an operational error on a mobile device.

Further inconsistencies are evident in the data shown in Figure~\ref{fig: DFRWS_boxplot_combi}. To ensure that the participants provided a response, the guidelines stipulated that no fewer than 0.5 stars could be awarded. As shown in Figure~\ref{fig: DFRWS_boxplot_combi}, participants nevertheless managed, albeit temporarily, to award zero stars. However, a subsequent review revealed that there were no deviations from the remaining guidelines for the questions and the error could not be replicated. We therefore assume that this was due to a temporary glitch in the Qualtrix software.

\subsection{Further Work}

As mentioned previously, the survey was originally designed to include a wider range of investigative authorities and is also limited by the fact that it was conducted exclusively within a single German state. However, it can serve as a basis for a larger survey with a more international scope, which could incorporate findings from different legal systems and a broader user base. Such a study could provide more detailed insight into the global relevance of usability in DF and take into account the issues and findings of this study.

Furthermore, this study provides valuable new insights for existing research in the field of DF, such as cooperation, training, and communication with investigators and stakeholders in criminal proceedings, or the development and improvement of KMS for DF.

Given the thematic breadth of the survey, which limited the number of questions per topic, future research could also further investigate the identified areas through more targeted qualitative or quantitative studies.

\section{Conclusion}
\label{sec:conclusion}

This study provides insight into how different stakeholders within the judicial system perceive the role and integration of digital forensics in criminal proceedings. Based on responses from 101 participants, including digital forensic experts, judges, prosecutors, investigators, executives, and administrative staff, the findings highlight the importance of collaboration, training, and organizational integration for the effective handling of digital evidence.

Participants strongly expressed their support for closer collaboration between digital forensic experts and other stakeholders, improved access to evidential data and results, and greater integration of topics related to digital forensics into stakeholder education and training. However, in fact, fewer than half of the participants reported having completed training in this area in the past five years. Although these findings indicate substantial potential for improvement in training and education, the perceived probability of an CSI-effect in digital forensics highlights the need for increased stakeholder awareness.

In addition, indications for large differences between public prosecution offices with internal digital forensic experts that performed significantly better than those without them in terms of skills and knowledge, handling cases involving digital evidence and overall satisfaction could be identified. Further differences were found between the participant groups, e.g., in triage assessments, and in requirements for knowledge management systems.

Overall, the study shows that many of the challenges faced in digital forensics extend beyond purely technical issues and are closely linked to organizational, human, and technical aspects of Usability within this area. These findings provide a foundation for future research on the practical integration and usability related aspects of digital forensics.

\bibliographystyle{ACM-Reference-Format}
\bibliography{bibliography}

\appendix

\section{Appendix -- An Analysis of Digital Forensics Research according to Usability Criteria}\label{sub: Related Issues in Digital Forensics}

For the development of the survey questions, we identified relevant previous work and mapped it to the three usability criteria: effectiveness, efficiency, satisfaction, based on their primary focus and reported results. Most of these studies were not originally framed within these categories and, therefore, may relate to multiple criteria. In the following, we present the selected additional works relevant to this study, organized according to effectiveness, efficiency, and satisfaction.

\subsection{Effectiveness}\label{sub: Examples of Effectiveness in DF}

Effectiveness can easily be transferred to the objectives of criminal proceedings. For example,  principles such as the integrity and authenticity of evidence, the application of established scientific methods or the traceability, reproducibility, and objectivity of results can be interpreted as measurements of quality, completeness, and precision in DF~\cite{Palmer.2001,Casey.2011,Dewald.2015}. 

With regard to the principles mentioned above, DF experts and investigators play a central role in ensuring the effectiveness of DF.~\citet{Casey.2011} and~\citet{Dewald.2015} emphasize their roles and the associated importance of thoroughly documenting their actions, procedures, and findings in the context of establishing the true facts of a case. The documentation and the expert's statements represent the embodiment of the results of DF for the court and must meet quality requirements as outlined above. 

Due to the important role law enforcement users play in the handling and interpretation of DF, their effective collaboration inevitably comes into focus.~\citet{Casey.2019} describe DF as a driver of decentralization within forensic sciences, allowing law enforcement agencies to establish their own dedicated teams. This decentralization, in turn, fosters closer collaboration between traditional investigators and DF experts.~\citet{Hansen.2017} report increased effectiveness resulting from this kind of interdisciplinary investigative work within police authorities. At the same time, they emphasize the necessity of mutual training to fully realize the potential of such collaboration. Furthermore,~\citet{Casey.2021} highlight the potential to improve effectiveness through shared knowledge management systems~(KMS).

In this sense,~\citet{Henseler.2018} also emphasize the need for training and continuing education for both legal professionals and DF experts in the fields of each other, in order to foster technical understanding and also raise awareness of legal requirements, thus enabling effective interdisciplinary collaboration between these fields.

However, there are differing views on whether DF can be conducted more effectively in decentralized or centralized environments.~\citet{Nouh.2019} emphasize the benefits of central coordination to increase effectiveness, particularly with regard to coordination, information exchange, and the avoidance of isolated knowledge silos. The Netherlands Forensics Institute~(NFI), as described by~\citet{Henseler.2018}, also serves as a central processing center for DF. According to their account, this approach offers advantages in terms of quality assurance, innovation capacity, market awareness, and the benefit of a single point of contact. The Hansken Digital-Forensics-as-a-Service~(DFaaS) System, developed by the NFI and described by~\citet{Beek.2020}, functions as a centralized DF software system, but it can also be deployed in decentralized configurations.

\subsection{Efficiency}\label{sub: Examples of Efficiency in DF}

DF is highly dependent on available resources, as reflected in persistent case backlogs identified as a critical issue in several studies~\citep{Casey.2019, Flory.2016, Hargreaves.2024, Horsman.2021}.~\citet{Pollitt.2013} characterizes this as an ongoing challenge in DF, particularly in time-sensitive cases and in the face of ongoing technological developments that require faster processing.

Additionally,~\citet{Baumgartner.2023} and~\citet{WilsonKovacs.2020} identify multiple bottlenecks that slow down DF investigations. Beyond legal and bureaucratic constraints, technological limitations and broader social factors contribute to backlogs. For example, while storage capacities continue to increase, data transfer speeds have not kept pace. The COVID-19 pandemic further intensified these problems by reducing the availability of DF personnel.

As described in Section~\ref{sub: Usability in Digital Forensics}, digital evidence triage aims to improve efficiency by reducing data volume.~\citet{Horsman.2022} distinguishes between in-lab and on-site triage. In-lab triage, conducted by experts, involves prioritizing or excluding devices and is largely reversible due to continued access to already secured evidence. In contrast, on-site triage is more critical, as decisions may permanently exclude potential evidence. This raises risks of false negatives, where relevant evidence is not secured or recognized, potentially leading to loss, compromised integrity, or irretrievability.

DF also requires significant financial and material resources. Both~\citet{Nouh.2019} and~\citet{WilsonKovacs.2020} note that budget cuts can negatively impact the efficiency and effectiveness of forensic work. Forensic software and hardware, along with the storage infrastructures and media necessary for backups, can incur substantial costs, particularly due to long retention periods~\citep{Baumgartner.2023}.

\subsection{Satisfaction}\label{sub: Examples of Satisfaction and UX in DF}

In contrast to effectiveness and efficiency, assessing user satisfaction in DF requires a broader perspective. These aspects are typically captured through user surveys, interviews, or observations, which are difficult to carry out in law enforcement~\citep{Flory.2016,Nouh.2019,Hargreaves.2024}. However, many relevant factors could be identified, though often only through closer examination. 

Perceptions and expectations of DF become especially apparent during the presentation of evidence in court, particularly testimony and documentation of methods and findings. Although there is little research on courtroom statements by DF experts,~\citet{Hargreaves.2024} found that most of the survey participants mentioned documentation as a common task, while courtroom testimony appeared to be much less frequent. Both are essential to establish trust in the objectivity, transparency and credibility of the applied methods and the evidence obtained~\cite{Casey.2011}.

\citet{Pollitt.2013} also notes that courts often expect faster processing of digital evidence than of analog materials.
\citet{WilsonKovacs.2020} highlight technical triage as a key example of expectations in DF. First responders are often expected to perform technical triage on their own after receiving only one training session and without routine. This can lead to reduced self-efficacy and increased reliance on DF experts, counteracting the intended efficiency gains. Errors from this stage are often attributed to DF, even without it's involvement, affecting it's perceived credibility. This can happen, e.g., when errors are detected or become apparent in later phases, such as processing or analysis, which are carried out by DF experts.

Another important aspect of perception and expectations in DF is cognitive bias, which refers to systematic involuntary deviations from reality that are difficult to avoid~\citep{Pohl.2022,Sunde.2019}.~\citet{Sunde.2019} describe these biases in DF as unconscious processes, influenced by emotions, that can reinforce existing expectations and compromise objectivity. Irrelevant case information and information flows can amplify these biases. They also note that countermeasures may not always be feasible, so these biases must be considered in the evidentiary process. In a later study~\citep{Sunde.2021b}, they found that DF experts were influenced by case information, with those receiving incriminating details finding more evidence of afflicting conditions while neglecting exonerating evidence.

Media portrayals of forensic science and the resulting, often unrealistic, perceptions and expectations in criminal proceedings, the so-called ``CSI effect''~\cite{Cole2009,Tyler.20250412}, are another example. The CSI effect was originally described in the context of court proceedings. Medial portrayals often create unrealistic expectations, such as generating high-quality images from heavily pixelated ones. This can become problematic for DF, for example, if those show effects are expected to be real and get demanded in court proceedings.

Despite overlaps with effectiveness and efficiency, integration and collaboration within the structures and organizations seem to be more closely linked to satisfaction.~\citet{Nouh.2019}, for example, point out that communication and information flows within law enforcement agencies differ significantly in terms of user experience compared to the private sector, which may hinder cooperation between investigators and DF units. Van Beek et al.~\citep{Beek.2020} and~\citet{Lawless.2022} also highlight that the implementation of DFaaS may require overcoming established organizational structures and practitioner cultures. In addition, such systems must be designed to include courts and defense lawyers.

Although DF is often associated with a high degree of individual autonomy and flexibility~\cite{Lawless.2022}, and increased collaboration with traditional investigators is frequently advocated~\citep{Beek.2020,Hansen.2017}, other sources reveal that in practice it is often constrained by rigid organizational structures, limited recognition, limited career prospects, and a work environment similar to assembly line labor~\citep{Buchheit.2020,Honekamp.2023,Merbach.2019,WilsonKovacs.2020}.

Role conflicts arising from such conditions can, as described by~\citet{Holt.2011}, trigger stress reactions among DF experts. According to~\citet{Kelty.2021}, these stress responses can range from general workload stress to burnout and secondary traumatic stress~(STS). They also found evidence that supportive management practices, strong leadership, and effective workplace environments can help mitigate stress. Additionally,~\citet{Strickland.2023} highlight that a lack of feedback on case progress and outcomes can reduce perceived appreciation, leading to lower motivation and increased frustration among DF experts and investigators.

\section{Appendix - Questions From the Survey}\label{app1}
This appendix provides the wording of the survey questions translated from German. Response options that were not selected (e.g., because only the judiciary participated) have been removed. Identifying information has been anonymized.

\subsection{Survey Introduction}

\emph{The introductory text of the survey has been translated and anonymized to prevent any identification of the authors in the review process.}
\\
\\
Thank you for your interest in our study!
\\\\
The objective of this study is to investigate perceptions and requirements related to digital forensics in criminal proceedings among different stakeholders of government involved in law enforcement and the judiciary. The study is inspired by research in usable security and human-centered security. Over the past decades, both fields have contributed substantially to making IT security solutions more effective, user-oriented, and better aligned with the needs of their target groups. Although digital forensics represents a subfield of IT security, it has so far only benefited to a limited extent from this research. In particular, there is a lack of studies that examine usability of digital forensics in multiple stakeholder groups. Therefore, this study focuses on the usability criteria of effectiveness, efficiency, satisfaction and user experience, as well as the differing perspectives between organizations and the selected current and future challenges in digital forensics.
\\\\
\underline{Purpose:} This scientific study examines aspects of usability and human-centered security and their applicability to digital forensics in the context of criminal proceedings. The focus is on courts, prosecution authorities, and law enforcement agencies within an [anonymized] German federal state.
\\\\
\underline{Requirements:} To participate in this study, you must:
\\
access the survey using a computer, tablet, or smartphone, and be at least 18 years old.
\\

\underline{Duration:} Participation is expected to take approximately \emph{20 minutes}. No risks beyond those associated with ordinary Internet use are anticipated. In particular, the survey does not contain questions about violence experiences or the mental health of participants.
Please complete all tasks with care and answer the questions as honestly as possible. Participation is voluntary and may be discontinued at any time before final submission of the questionnaire. After interruption, participants may continue the survey within one week, after which incomplete responses will be discarded.
\\\\
\underline{Contact:} This study is conducted as part of an academic research project at an [anonymized] German university. If you have any questions or encounter technical problems, you may contact the research team via the email address provided in the survey materials.
\\\\
\underline{Data Protection:} Responses are stored under a randomly generated participation ID. Data are collected and processed for scientific research purposes. Where technically necessary, data may be transferred to third party service providers involved in survey operation or data storage; such cases are indicated within the survey. By starting the questionnaire, participants consent to the processing of their data for the purposes of this study. Participants may withdraw their consent at any time and may request access to, correction to, restriction of processing, or deletion of their stored data by contacting the research team and providing their participation ID.
\\\\
\emph{Your participation ID: <generated ID>}
\\\\
Please store this ID together with the contact information provided in the survey materials, as it may be required for future inquiries.

\subsection{Participation Consent}

Do you agree to the collection and processing of your data?\\ 
$\circ$ Yes\\ 
$\circ$ No

\subsection{Demographic Questions}

Please indicate your age group:\\
$\circ$ 18 to 29\\
$\circ$ 30 to 39\\
$\circ$ 40 to 49\\
$\circ$ 50 to 59\\
$\circ$ 60 to 69\\
$\circ$ I prefer not to say\\

Please indicate the group or institution to which you belong:\\
$\circ$ Justice - Higher Regional Court, Regional Court, or Local Court\\
$\circ$ Justice - Civil, Labor, or Administrative Court\footnote{This response option was included as a filter to ensure that only those judges who actually handle criminal cases were included. (not chosen)}\\
$\circ$ Justice - Public Prosecutor’s Office\\
$\circ$ Justice - Administration/ IT\\
$\circ$ Police - Emergency Response and Operations or Traffic Police\\
$\circ$ Police - Criminal Investigation Division\\
$\circ$ Police - Administration/ IT\\

Please select your type of employment. If none of the options apply, please select ‘Other employment’.\\
$\circ$ Judge\\
$\circ$ Public Prosecutor\\
$\circ$ Digital Forensics Expert\\ 
$\circ$ Investigator\\
$\circ$ Executive Position\\
$\circ$ Administration / IT\\ 
$\circ$ Other employment\\

Please indicate your employment status:\footnote{Other options (``Freelancer'' and ``Apprentice, student, or legal trainee'') not chosen.}\\
$\circ$ Employee\\
$\circ$ Civil servant\\

Can you provide information about your pay grade or employment group?\\
$\circ$ Intermediate service\\
$\circ$ Upper intermediate service\\ 
$\circ$ Higher service\\
$\circ$ Other classification or pay scale\\ 
$\circ$ I prefer not to say\\

Please indicate which of the following best describes your highest level of education:\\
$\circ$ No vocational qualification\\
$\circ$ Vocational training\\
$\circ$ Advanced professional qualification (e.g., Certified Specialist)\\
$\circ$ Undergraduate degree, e.g., Bachelor/ Dipl.~(FH)\\
$\circ$ Graduate degree – Master/ Dipl. (University)/ First or Second State Exam/ Ph.D.\\

\subsection{Training and Access to Expertise}

Have you attended training courses related to handling digital evidence, digital forensics, or IT security/ cybercrime within the last 5 years?\\$\circ$ Yes\\
$\circ$ No\\

What aspects were covered in these training courses?
\\$\square$ Digital forensics\\
$\square$ Digital evidence\\
$\square$ Technical aspects of cybercrime/IT security\\
$\square$ Legal aspects including IT law\\
$\square$ Criminalistic aspects\\
$\square$ Phenomenology/ manifestations\\

Does your agency or organization employ experts in the field of digital forensics?\\
$\circ$ Yes\\
$\circ$ No\\
$\circ$ I don’t know\\

If known, indicate the educational or professional background of the digital forensics or cybercrime specialists employed by your agency/organization.\footnote{The question was originally intended to distinguish between a broader range of participants (e.g., the police). It was omitted from the paper.}\\
$\circ$ Person(s) with IT/ technical or scientific qualifications;\\
$\circ$ Person(s) with legal or economic qualifications;\\
$\circ$ Person(s) with both qualifications; $\circ$ Qualifications unknown\\

How many experts in digital forensics/cybercrime are employed by your agency or organization?\footnote{Other options not chosen}\\
$\circ$ 1\\
$\circ$ 2–3\\
$\circ$ 4–6\\
$\circ$ I don’t know\\

Are you aware of any reasons why your agency or organization employs no or only a few experts in digital forensics/cybercrime?\\
$\square$ No need\\
$\square$ Sufficient availability of external experts\\
$\square$ Lack of funding/budget for the hiring of experts\\
$\square$ Lack of suitable applications for open positions\\
$\square$ Sufficient internal expertise\\
$\square$ Other, not listed reason\\

\subsection{Collaboration and Digital Skills}

\emph{The question below addresses collaboration and competencies for the following authorities or organizations: \\
``Courts'', 
``PPOs'', 
``Focal PPOs'', 
``Police'', 
``Tax Investigation'', 
``Academic Sector'', 
and ``Private Business Sector''. 
\\
The question also serves to filter which answer options are displayed in the following items of this section.}\\

Please indicate whether you collaborate with the following agencies and, if so, what level of digital competencies they possess in criminal proceedings. If your own agency or organization is listed, please refer to collaboration with comparable institutions.\\
For each of the groups listed below:\\
$\circ$ No collaboration\\
$\circ$ Collaboration – no specific digital competencies are known\\
$\circ$ Collaboration – competencies in handling digital evidence\\
$\circ$ Collaboration – competencies in digital forensics present\\

\emph{The following four questions each have nine possible answers—represented as star ratings—for the organizations listed in parentheses (``own agency/organization'', ``Courts'', ``PPOs'', ``focal PPOs'', ``Police'', ``Tax Investigation'', ``Academic Sector'', and ``Private Business Sector''). With the exception of one’s own agency, cooperation with the respective organizations must have been indicated in the previous question.}

\begin{packed_enum}
\renewcommand{\labelenumi}{\arabic{enumi})}
\item Rate the abilities of the following groups to understand and apply digital evidence or digital forensic methods.\\

\item How would you assess the ability of the following authorities or organizations to handle a case involving digital evidence effectively?\\

\item How would you evaluate the financial and material resources of the listed authorities/organizations with regard to meeting future challenges such as increasing data volumes and other digital issues in criminal prosecution?\\

\item Evaluate your general satisfaction with the collaboration with the organizations listed below, regardless of your previous responses. This question refers exclusively to overall satisfaction and not specifically to digital forensics or digital evidence.
\end{packed_enum}

\subsection{Operational Challenges of Usability in DF}

\emph{All questions in this group are modeled as 5-point Likert-scale with 1 = stongly disagree, ..., 5 = strongly agree; most questions below have the prefix "Do you agree that [...]", which is omitted for brevity. The order of questions is randomized.}\\

\begin{packed_item}

    \item ... the backlog of digital evidence is a major problem in the investigation or criminal procedure?

    \item ... scientific methodologies, procedural models, and established practices are necessary for the effective conduct of digital forensics?

    \item ... digital evidence in criminal proceedings should be introduced as a separate form of evidence alongside witness, expert, documentary, and visual evidence to adequately consider technological progress, ensure the integrity of evidence, minimize evidential uncertainties, and facilitate compliance with international standards?

    \item  ... a closer connection between traditional investigative work and digital forensics in the form of interdisciplinary investigative teams has the potential to increase effectiveness, efficiency, and satisfaction with investigative work?

    \item ... expanding access to forensic software through the use of web interfaces or virtual desktops in traditional investigative workplaces or courtrooms could improve the usability of digital forensics?

    \item ... the discovery and securing of potential digital evidence is one of the main tasks of search teams and first responders.

    \item ... decentralized processing of digital forensics on-site may be more effective, efficient, and satisfying than centralized processing in regional centers?

    \item ... it would be useful to integrate digital evidence and digital forensics more into the training and education of police, tax investigators, and lawyers?

    \item Do you think that improved processes for the transfer of information and digital evidence to digital forensics could increase the efficiency of their work?

    \item ... ergonomic office equipment, adequate climate control, a quiet work environment,and retreat spaces are necessary for effective workplaces in digital forensics and the analysis of their results?

    \item Do you think that regular feedback from law enforcement agencies and courts on digital forensics could help better adapt to the needs of their target groups and gain increased perceived appreciation?

    \item Do you see a benefit in first responders and search teams using an app to assist them in recognizing, capturing, and handling digital evidence in a forensically correct manner?

    \item Do you believe that the simplified or exaggerated portrayal of forensic methods in the media or popular crime series also influences the perception and expectations of digital forensics?

    \item ... the preselection (triage) of digital evidence based on certain criteria during a search operation can be an effective means of reducing data volumes and increasing efficiency in investigations?

    \item ... the preliminary inspection of storage devices using software to identify suspicious data during a search operation is an effective means of identifying relevant data and excluding irrelevant storage devices?
    
\end{packed_item}

\subsection{Knowledge Management}

Do you see a need for an inter-agency knowledge management portal that allows various professional groups to exchange information, educate themselves, and stay informed about digital forensics, evidence, and current methods and developments?\\
$\circ$ Yes, I see a need\\
$\circ$ No, I do not see a need\\

What functions and aspects are you particularly interested in for a knowledge management portal in the field of digital forensics and digital evidence? Please choose from the following options.\\
$\square$ Restricted areas for teams and agencies\\
$\square$ Knowledge databases for basic knowledge\\
$\square$ Guides and how-to manuals\\
$\square$ Digital certifications and rating functions\\
$\square$ Search function for experts and contacts\\
$\square$ Exchange platforms for software and source code\\
$\square$ Involvement of academia \\
$\square$ Technical knowledge database for experts in digital forensics\\
$\square$ Business and industry participation \\
$\square$ Cross-border usability\\

What aspects would you focus on in a knowledge management portal?\\
$\square$ Information on developments and innovations\\
$\square$ Access to information on specific knowledge areas\\
$\square$ Exchange with people in your field\\
$\square$ Exchange with people from different fields\\
$\square$ Opportunities for self-education\\
$\square$ Finding contacts\\
$\square$ Access to guidelines and recommendations\\

\subsection{Retaining or Recruiting DF Experts}

How would you assess the difficulty of recruiting in your agency?\\
\emph{Tabular query for the points ``Digital Forensics'' and ``Agency overall.''}\\
$\circ$ Very difficult \\
$\circ$ Quite difficult \\
$\circ$ Neither easy nor difficult\\
$\circ$ Quite easy\\
$\circ$ Very easy\\

What measures do you consider suitable for attracting new employees to the field of digital forensics or retaining existing personnel?\\
$\square$ Offering a degree program with subsequent commitment\\
$\square$ Promotion to higher positions\\
$\square$ Allowing for specialized careers\\
$\square$ Finance of further education/studies\\
$\square$ Higher base salary or higher entry levels\\
$\square$ Collaboration with academia and research\\
$\square$ Option for the status of a civil servant\\
$\square$ Pre-granting of experience levels\\
$\square$ Allowances (e.g., specialist allowances, etc.)\\

Are you willing to answer two additional questions regarding the relationship between IT security and law enforcement or justice?\\
$\circ$ Yes\\
$\circ$ No\\

\emph{[if yes ]} Individuals and researchers in the field of IT security often use offensive methods (colloquially hacking) to systematically test and research the effectiveness of security measures. There is a risk that these methods, due to their similarity to the actions of criminal hackers and the insufficient distinction in legislation (e.g. §§ 202a, 202c StGB), could lead to conflicts with law enforcement. Do you think these circumstances could contribute to greater distance and skepticism from people in the IT sector towards law enforcement?\\
$\circ$ Definitely yes\\
$\circ$ Probably yes\\
$\circ$ Probably not\\
$\circ$ Definitely not\\
$\circ$ I do not want to answer this question\\

Do you believe, conversely, that there is skepticism towards computer science and IT security from law enforcement and the judiciary, as the differences between actions in the field of IT security are often equated with criminal actions, or the differences are not perceived or not sufficiently legally differentiated?\\
$\circ$ Definitely yes\\
$\circ$ Probably yes\\
$\circ$ Probably not\\
$\circ$ Definitely not\\
$\circ$ I do not want to answer this question\\

\subsection{Concluding Question}

Do you think that the better improved usability of the processes, products, and results of digital forensics will become more important in the future, especially in criminal proceedings?\\
$\circ$ Definitely yes\\
$\circ$ Probably yes\\
$\circ$ Probably not\\ 
$\circ$ Definitely not\\
$\circ$ I don’t know\\

\subsection{Attention Question}

Thank you for participating in our survey. Your opinion is extremely valuable to our research.
\\
We want to ensure that our results are as accurate as possible. Please let us know if you completed the survey carefully or if you participated out of curiosity or for other reasons (which is also okay for us, but we ask you to indicate this so that data quality remains high).
\\
We appreciate your openness and assure you that your responses will be treated confidentially.\\
$\circ$ Yes, I completed the survey carefully. Data should be included in the study.\\
$\circ$ No, I did not complete the survey with care becuase of curiosity or other reasons. Data should not be included in the study.

\end{document}